\pdfoutput=1
\newcommand*{\ATLASLATEXPATH}{latex/}
\documentclass[cernpreprint,texlive=2016,UKenglish]{\ATLASLATEXPATH atlasdoc}

\usepackage{\ATLASLATEXPATH atlaspackage}
\usepackage{\ATLASLATEXPATH atlasbiblatex}
\usepackage[process]{\ATLASLATEXPATH atlasphysics}
\usepackage{HION-2018-02-PAPER-defs}

\usepackage{dcolumn}

\graphicspath{{logos/}{figures/}}

\AtlasTitle{Measurements of $W$ and $Z$ boson production in $pp$ collisions at $\sqrt{s}=5.02$~\TeV\ with the ATLAS detector}

\author{The ATLAS Collaboration}

\AtlasRefCode{HION-2018-02}

\PreprintIdNumber{CERN-EP-2018-259}

\AtlasDate{\today}

\AtlasJournal{EPJC}

\AtlasAbstract{%
  Measurements of fiducial integrated and differential cross sections for inclusive $W^+$, $W^-$ and $Z$ boson production are reported.
  They are based on $25.0 \pm 0.5\; \mathrm{pb}^{-1}$ of $pp$ collision data at $\sqrt{s} = 5.02$~\TeV\ collected
  with the ATLAS detector at the CERN Large Hadron Collider.
  Electron and muon decay channels are analysed, and the combined $W^+$, $W^-$ and $Z$ integrated cross sections are found to be
  $\sigmaWp = 2266 \pm 9~\text{(stat)} \pm 29~\text{(syst)} \pm 43~\text{(lumi)}$~pb,
  $\sigmaWm = 1401 \pm 7~\text{(stat)} \pm 18~\text{(syst)} \pm 27~\text{(lumi)}$~pb,
  and $\sigmaZ = 374.5 \pm 3.4~\text{(stat)} \pm 3.6~\text{(syst)} \pm 7.0~\text{(lumi)}$~pb,
  in good agreement with next-to-next-to-leading-order QCD cross-section calculations.
  These measurements serve as references for Pb+Pb interactions at the LHC at $\sqrt{s_{\mathrm{NN}}}=5.02$~\TeV.
}

\hypersetup{pdftitle={ATLAS draft},pdfauthor={The ATLAS Collaboration}} 

\begin{document}

\maketitle

\section{Introduction \label{sec:intro}}
Measurements of \Wpm\ and \Z\ boson\footnote{Throughout this paper, $\Zg$ boson production is referred to as $Z$ boson production.} 
production at hadron colliders provide a benchmark for the understanding of quantum 
chromodynamics~(QCD) and electroweak~(EW) processes. Predictions 
for the differential and fiducial cross sections
are available up to next-to-next-to-leading-order~(NNLO) 
accuracy in QCD and include EW corrections at next-to-leading-order~(NLO) 
accuracy~\cite{Anastasiou:2003ds,Catani:2007vq,Catani:2009sm}. The rapidity 
distribution of EW boson production is sensitive to the underlying QCD dynamics and, in particular, to the parton 
distribution functions (PDFs) which define the initial kinematics of the hard process. Therefore, measurements of weak-boson 
production offer an excellent opportunity to test models of parton dynamics.

The ATLAS, CMS and LHCb collaborations have measured \Wpm\ and \Z\ boson production in proton--proton~(\pp) collisions at centre-of-mass energies of $\sqrt{s} = 7$, 8 and 13~\TeV~\cite{Aaboud:2016btc,Aad:2016naf,Chatrchyan:2014mua,Aaij:2015zlq}. These measurements 
provide precision tests of the QCD theory and PDFs, which can be complemented with measurements at the additional centre-of-mass energy $\sqrt{s} = 5.02$~\TeV.

This paper describes measurements of the production cross sections times leptonic branching ratios for the inclusive \wplnu, \wmlnu\ and \zll~$(\ell = e,~\mu)$ processes. 
Integrated and differential cross sections are measured in a fiducial phase space defined by detector acceptance and lepton kinematics.
For \Wpm\ bosons the decay lepton charge asymmetry is also determined.
All measurements are performed with \pp collision data corresponding to an integrated luminosity of $25.0~\mathrm{pb^{-1}}$, 
collected at $\sqrt{s} = 5.02$~\TeV\ with the ATLAS detector. The data were recorded during the autumn of 2015. The peak 
instantaneous luminosity delivered by the LHC was $L = 3.8 \times 10^{32}$~$\mathrm{cm^{-2}\mathrm{s^{-1}}}$ and the mean number of \pp 
interactions per bunch crossing (hard scattering and pile-up events) was 1.5. Therefore, this dataset is characterised 
by a relatively low pile-up contribution as compared to the measurements of weak-boson production performed at higher 
centre-of-mass energies by ATLAS.

In addition, the measurement of \Wpm\ and \Z\ boson production in \pp collisions at the centre-of-mass energy $\sqrt{s} = 5.02$~\TeV\ is an important reference for weak-boson production in heavy-ion collisions. The LHC has provided both proton--lead~($p$+Pb) and lead--lead~(Pb+Pb) collisions at the centre-of-mass energy per nucleon pair $\sqrt{s_{\mathrm{NN}}}=5.02$~\TeV. 
Published results from the ATLAS and CMS collaborations are currently available 
for \Wpm\ and \Z\ boson production~\cite{Aad:2014bha,Aad:2012ew,Chatrchyan:2012nt,Chatrchyan:2014csa} in Pb+Pb collisions at $\sqrt{s_{\mathrm{NN}}}=2.76$~\TeV\ and \Z\ boson production~\cite{Aad:2015gta, Khachatryan:2015pzs} in the $p$+Pb system at $\sqrt{s_{\mathrm{NN}}}=5.02$~\TeV.

\section{The ATLAS detector\label{sec:atlas}}

The ATLAS experiment~\cite{PERF-2007-01} is a multipurpose particle detector with a forward--backward symmetric 
cylindrical geometry.\footnote{ATLAS uses a right-handed coordinate system with its origin at the nominal interaction point (IP) 
in the centre of the detector and the $z$-axis along the beam pipe. The $x$-axis points from the IP to the centre of the LHC ring, 
and the $y$-axis points upward. Cylindrical coordinates $(r,\phi)$ are used in the transverse plane, $\phi$ being the azimuthal angle 
around the $z$-axis. The pseudorapidity is defined in terms of the polar angle $\theta$ as $\eta=-\ln\tan(\theta/2)$,
while the rapidity is defined as $y = \frac{1}{2} \ln{\frac{E+p_z}{E-p_z}}$.}
It consists of an inner tracking detector surrounded by a thin superconducting solenoid, electromagnetic and hadronic calorimeters,
and a muon spectrometer incorporating three large superconducting air-core toroid magnets with eight coils each.

The inner-detector system (ID) is immersed in a \SI{2}{\tesla} axial magnetic field 
and provides charged-particle tracking in the pseudorapidity range $|\eta| < 2.5$. At
small radii, a high-granularity silicon pixel detector covers the
interaction region and typically provides four measurements per track. It is followed by the silicon 
microstrip tracker, which usually provides
eight measurement points per track. These silicon
detectors are complemented by a gas-filled straw-tube transition
radiation tracker, which enables track reconstruction up to $|\eta| = 2.0$. 
The transition radiation tracker also provides electron
identification information based on the fraction of hits (out of $\sim 35$
in total) with an energy deposit above a threshold indicative of
transition radiation. 

The calorimeter system covers the pseudorapidity range $|\eta| < 4.9$.
Within the region $|\eta|< 3.2$, electromagnetic (EM) calorimetry is
provided by high-granularity lead/liquid-argon (LAr) calorimeters,
with an additional thin LAr presampler covering $|\eta|<1.8$  
to correct for upstream energy-loss fluctuations. The EM calorimeter
is divided into a barrel section covering $|\eta|<1.475$ and two
endcap sections covering $1.375<|\eta|<3.2$. For $|\eta|<2.5$ it is
divided into three layers in depth, which are finely segmented in
$\eta$ and $\phi$. Hadronic calorimetry is provided by a
steel/scintillator-tile calorimeter, segmented into three barrel
structures within $|\eta| < 1.7$ and two copper/LAr hadronic endcap
calorimeters covering $1.5<|\eta|<3.2$. The solid-angle coverage is completed with forward
copper/LAr and tungsten/LAr calorimeter modules in $3.1<|\eta|<4.9$, optimised for
electromagnetic and hadronic measurements, respectively. 

The muon spectrometer (MS) comprises separate trigger and
high-precision tracking chambers measuring the deflection of muons in
the magnetic field generated by the toroid magnets. The
precision chamber system covers the region $|\eta| < 2.7$ with three
layers of monitored drift tubes, complemented by cathode strip
chambers in the forward region. The
muon trigger system covers the range $|\eta| < 2.4$ with resistive
plate chambers in the barrel, and thin gap chambers in the endcap
regions. 

In 2015, the ATLAS detector had a two-level trigger system~\cite{Aaboud:2016leb}. The level-1 trigger is implemented in
hardware and uses a subset of detector information to reduce the event
rate to a value of at most \SI{75}{\kHz}. This is followed by
a software-based high-level trigger
 which reduces the event rate to about \SI{1}{\kHz}.

\section{Simulated event samples \label{sec:samples}}

Samples of Monte Carlo~(MC) simulated events are used to evaluate the selection efficiency for signal events and the contribution of several background processes to the analysed dataset. All of the samples are processed with the
\textsc{Geant4}-based simulation~\cite{Agostinelli:2002hh, SOFT-2010-01} of the ATLAS detector. Dedicated efficiency and calibration 
studies with data are used to derive correction factors to
account for residual differences between experiment and simulation, as is subsequently described.

The processes of interest, specifically events containing \Wpm or \Z\ bosons, were generated with the \Powheg-\textsc{Box} v2 MC program~\cite{Alioli:2010xd} interfaced to the \Pythia\ 8.186 parton shower model~\cite{Sjostrand:2007gs}. The CT10 PDF set~\cite{Gao:2013xoa} was used in the matrix element, while the CTEQ6L1 PDF set~\cite{Pumplin:2002vw} was used with the AZNLO~\cite{STDM-2012-23} set of generator-parameter values (tune) for the modelling of non-perturbative effects in the initial-state parton shower. The \Photos++ v3.52 program~\cite{Davidson:2010ew} was used for QED radiation 
from electroweak vertices and charged leptons. Samples of top-quark pair~(\ttbar) and single-top-quark production were generated with the \Powheg-\textsc{Box} v2 generator, which uses NLO matrix element calculations together with the CT10f4 PDF set~\cite{Lai:2010vv}. Top-quark spin correlations were preserved for all top-quark processes. The parton shower, fragmentation, and underlying event were simulated using \Pythia\ 6.428~\cite{Sjostrand:2006za} with the CTEQ6L1 PDF set and the corresponding Perugia 2012 tune (P2012)~\cite{Skands:2010ak}. The top-quark mass was set to 172.5~\GeV. The EvtGen v1.2.0 program~\cite{Lange:2001uf} was used to model
bottom and charm hadron decays for all versions of \Pythia. Diboson processes were simulated using the \Sherpa\ v2.1.1 generator~\cite{Gleisberg:2008ta}. They were calculated for up to one~($ZZ$) or zero ($WW$, $WZ$) additional partons at NLO QCD accuracy and up to three additional partons at LO. In addition, the \Sherpa\ diboson sample cross section is scaled to account for the cross section change when the $G_\mu$ scheme~\cite{Hollik:1988ii} is used instead of the native one
for the EW parameters, resulting in an effective value of $\alpha \approx 1/132$. Multiple overlaid \pp collisions were simulated with the soft QCD processes of \Pythia\ v8.186
using the A2 tune~\cite{ATL-PHYS-PUB-2012-003} and the MSTW2008LO PDF set~\cite{Martin:2009iq}. For the comparison with data in 
differential distributions and 
the evaluation of single-boson EW backgrounds for the cross-section calculations, the single-boson simulations are normalised to the results of NNLO QCD calculations
obtained with a modified version of DYNNLO 1.5~\cite{Catani:2007vq,Catani:2009sm} optimised for speed of computation, with uncertainties of 3\%. 
The simulations of all other processes are 
normalised to the predictions of NLO QCD calculations, with uncertainties of 10\% for the diboson and top-quark processes.

\section{Object definitions and event selection \label{sec:selection}}

This section describes the reconstruction of electrons, muons and hadronic recoil objects, and the selection of $W$ and $Z$ bosons. Candidate events are required to have at least one primary vertex reconstructed from at least three tracks with $\pt$ > 400~MeV and to pass a trigger selection, which requires a single electron or muon candidate with a $\pt$ threshold of 15~\GeV\ or 14~\GeV, respectively. In addition, a loose likelihood-based identification requirement~\cite{Aaboud:2016vfy, ATLAS:2016iqc} is applied in the electron trigger.

Electron candidates are required to have $\pT>20~(25)$~\GeV\ in the $Z$ ($W$) boson analysis and $|\eta|$ < 2.47. Candidates within the transition region between barrel and endcap calorimeters (1.37 < $|\eta|$ < 1.52) are rejected.
In addition, medium likelihood-based identification and tight isolation requirements are applied~\cite{Aaboud:2016vfy, ATLAS:2016iqc}. 
Muon candidates must satisfy $\pT>20~(25)$~\GeV\ in the $Z$ ($W$) boson analysis and $|\eta|$ < 2.4
and pass the requirements of medium identification and 
tight isolation~\cite{Aad:2016jkr}; both criteria were optimised for 2015 analysis conditions.

Additional requirements are imposed on the significance of the transverse impact parameter, $d_0$, such that $|d_0|/\sigma_{d_0}<5~(3)$ for electron~(muon) candidates. 
To ensure that lepton candidates originate from the primary vertex, a requirement is also placed on the longitudinal impact 
parameter, $z_0$, multiplied by the sine of the track polar angle, $\theta$, such that the absolute value is smaller than $0.5$~mm.

Events with $Z$ boson candidates are selected by requiring exactly two opposite-charge electrons or muons, at least one of which is matched to a lepton selected at trigger level.
The dilepton invariant mass must satisfy the fiducial requirement $66<\mll<116$~\GeV.

Events with $W$ boson candidates are selected by requiring exactly one electron or muon that is matched to a lepton selected at trigger level.
The (anti-)neutrinos from \wlnu\ decays escape direct detection. A measure of the neutrino transverse momentum, \ptnu, can be inferred from information about the hadronic system recoiling against the $W$ boson.
The hadronic recoil is the vector sum of all calorimeter energy clusters excluding the deposits from the decay muon or electron, and is further described below. 
The transverse projection of the recoil onto the $r$--$\phi$ plane, \utvec, is used together with the decay lepton transverse momentum \ptlvec\ for the calculation of the missing transverse momentum vector,
\begin{eqnarray}
\nonumber
\label{eq:met} \metvec &=& - \left( \utvec + \ptlvec \right),
\end{eqnarray}
whose magnitude is denoted \met.
The transverse mass of the lepton-\met\ system is defined as $\mt = \sqrt{2 \ptl \met \left(1-\cos{\Delta\phi_{\ell,\met}}\right)}$ where $\Delta\phi_{\ell,\met}$ is the azimuthal angle between \ptlvec\ and \metvec.
The $W$~boson candidate events are selected by requiring $\met > 25$~\GeV\ and $\mt > 40$~\GeV. These event selection requirements are optimised to reduce background contributions from multi-jet processes.

The general structure of the algorithm used for hadronic recoil reconstruction is introduced in Ref.~\cite{Aaboud:2017svj}, where three-dimensional topological clusters~\cite{Aad:2016upy} calibrated at the hadronic scale are used as inputs to the algorithm. In this measurement, the hadronic recoil is reconstructed using particle flow objects~\cite{Aaboud:2017aca} as inputs. The ATLAS particle flow algorithm provides an improved \met\ resolution compared to the algorithm using only topological clusters, and makes the measurement less sensitive to pile-up by separating the 
charged-hadron contribution from the neutral hadronic activity~\cite{Aaboud:2017aca}. The charged activity is measured by 
the ID and the related tracks from charged hadrons can be matched to a vertex. From all charged hadrons, only calorimetric clusters associated with a track originating from
the reconstructed primary vertex are retained as input to the hadronic recoil algorithm. The neutral hadronic activity is represented by clusters without an associated track, and is also used in the recoil algorithm.

\section{Detector performance corrections \label{sec:perf}}

\subsection{Lepton calibration and efficiency}

The electron energy calibration is primarily obtained from the simulation by employing multivariate techniques~\cite{Aad:2014nim}.
The signal $Z\rightarrow ee$ MC simulation is used for deriving the data energy scale calibration and resolution corrections for the simulation. The energy resolution is corrected with additional factors no larger than about 1\% in the barrel and up to 2\% in the endcap region of the detector in order to account for a slightly worse resolution observed in the data. The energy scale 
is corrected by applying a per-electron energy scale factor to the data derived from a comparison of the electron-pair invariant mass 
between the simulation and the data. This procedure was found to be sensitive to the pile-up distribution in data due to different settings used for the signal readout from the EM calorimeters~\cite{Abreu:2010zzc}. Therefore, a special set of scale correction factors was derived for this dataset.

Measurements of muon momenta can be biased by the detector alignment and resolution, distortions of the magnetic 
field or imprecise estimates of the amount of passive material in the detector. Corrections of the muon momentum scale and resolution, which are 
applied to the simulation, are derived as a function of the muon $\eta$ and $\phi$ using \zmm\ events~\cite{Aad:2016jkr}. 
The correction factors are chosen such that they minimise the $\chi^2$ between the muon-pair invariant mass distributions 
in data and simulation.

Electron candidates used for the analysis are required to satisfy selection criteria related to 
reconstruction, identification, isolation and trigger. For each of these requirements, the efficiency of the selection 
is measured in data with the tag-and-probe method in \zee\ events, as described in Ref.~\cite{ATLAS:2016iqc}, and compared with the simulation. Data-to-simulation ratios of efficiencies are
used as scale factors to correct the simulation for the observed differences. Measurements are performed as a function of the electron \pt\ and $\eta$ for electrons selected in the analysis. All uncertainties related to efficiency are classified as either correlated or uncorrelated, and are propagated accordingly to the final measurement uncertainty.

The electron reconstruction efficiency is in the range 95--99\% both in the data and simulation and is typically 
measured with a precision of 2\%. The data-to-simulation ratio is up to 2\%~(5\%) different from unity in the barrel~(endcap) 
calorimeter and is measured typically with 2\% precision for \pt\ in the range $\sim$30--50~\GeV\ and 
5\% for $\pt>60$~\GeV. The efficiency of an electron to further pass the medium identification definition varies from 85\% to 95\% and is measured with 2\% precision. This efficiency differs from the efficiency measured in the MC simulation by up to 5\%. The isolation efficiency is measured with a precision of 5\% and agrees with the simulated value within 
2\%. Data-to-simulation correction factors for identification and isolation efficiencies are measured with a precision of 2--6\%. Finally, the trigger efficiency data-to-simulation ratio is found to deviate from unity by 0.5--3\%\ and is measured with a precision of up to 2\%.

Various selection requirements related to muon trigger, reconstruction, 
identification and isolation are imposed on muon candidates used in the analysis. The efficiency of the selection criteria is 
measured in data with the tag-and-probe method in \zmm\ events~\cite{Aaboud:2016leb,Aad:2016jkr} and compared with the simulation. 
Ratios of the efficiencies determined in data and simulation are applied as scale factors to correct the simulated events. 
For muons with $\pt>20$~\GeV, the correction factors measured as a function of muon \pt\ have typically an uncertainty of 1--2\% and do not deviate from a constant value by more than 3\%. Therefore, the \pt\ dependence of the scale factors is neglected, and they are evaluated only as a function of muon $\eta$.

The muon trigger efficiency in the endcap region of the detector~($1.05<|\eta|<2.4$) is measured to be around 90\%, and 
the values obtained in data and simulation agree well. However, in the barrel region~($|\eta|<1.05$) the trigger efficiency 
determined in the simulation varies from 70\% to 85\%, while the efficiency measured in data is lower by 5--15\%, which 
results in sizeable scale factors. The combined reconstruction and identification efficiency for medium-quality muons typically 
exceeds 99\% in both the data and simulation with good agreement between the two measurements.
The efficiency of the isolation selection is found to be 97--98\% in the MC simulation and it differs from the efficiency measured in the data by about 2\% in the most central~($|\eta|<0.6$) and most forward detector regions~($1.74<|\eta|<2.4$).

All measurements of lepton efficiency corrections are limited in their precision by the number of \zll\ candidates available in the $\sqrt{s} = 5.02$~\TeV\ dataset.

\begin{figure}[!t]
  \centering
  \subfloat[]{\includegraphics[width=0.49\textwidth]{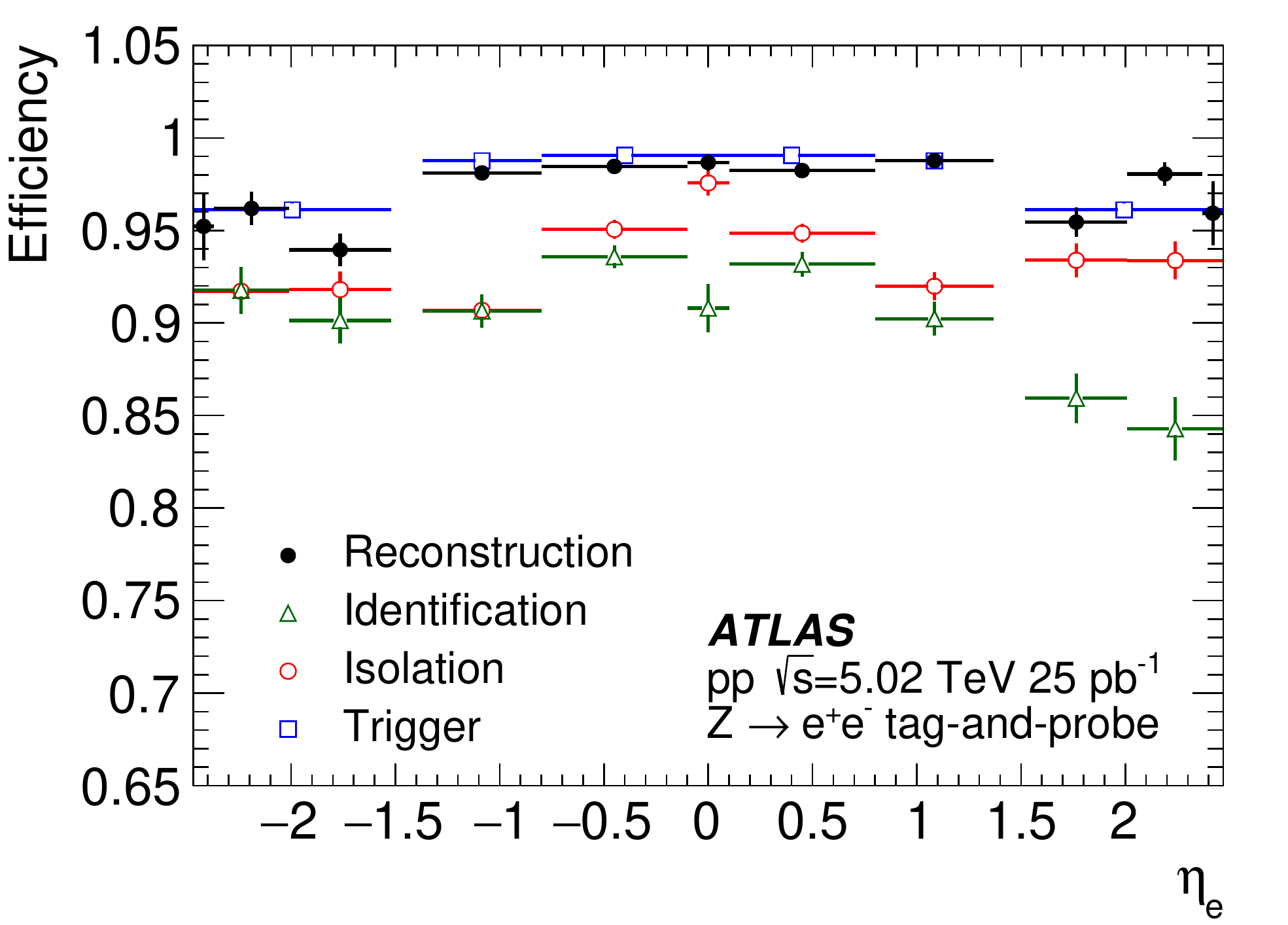}}
  \subfloat[]{\includegraphics[width=0.49\textwidth]{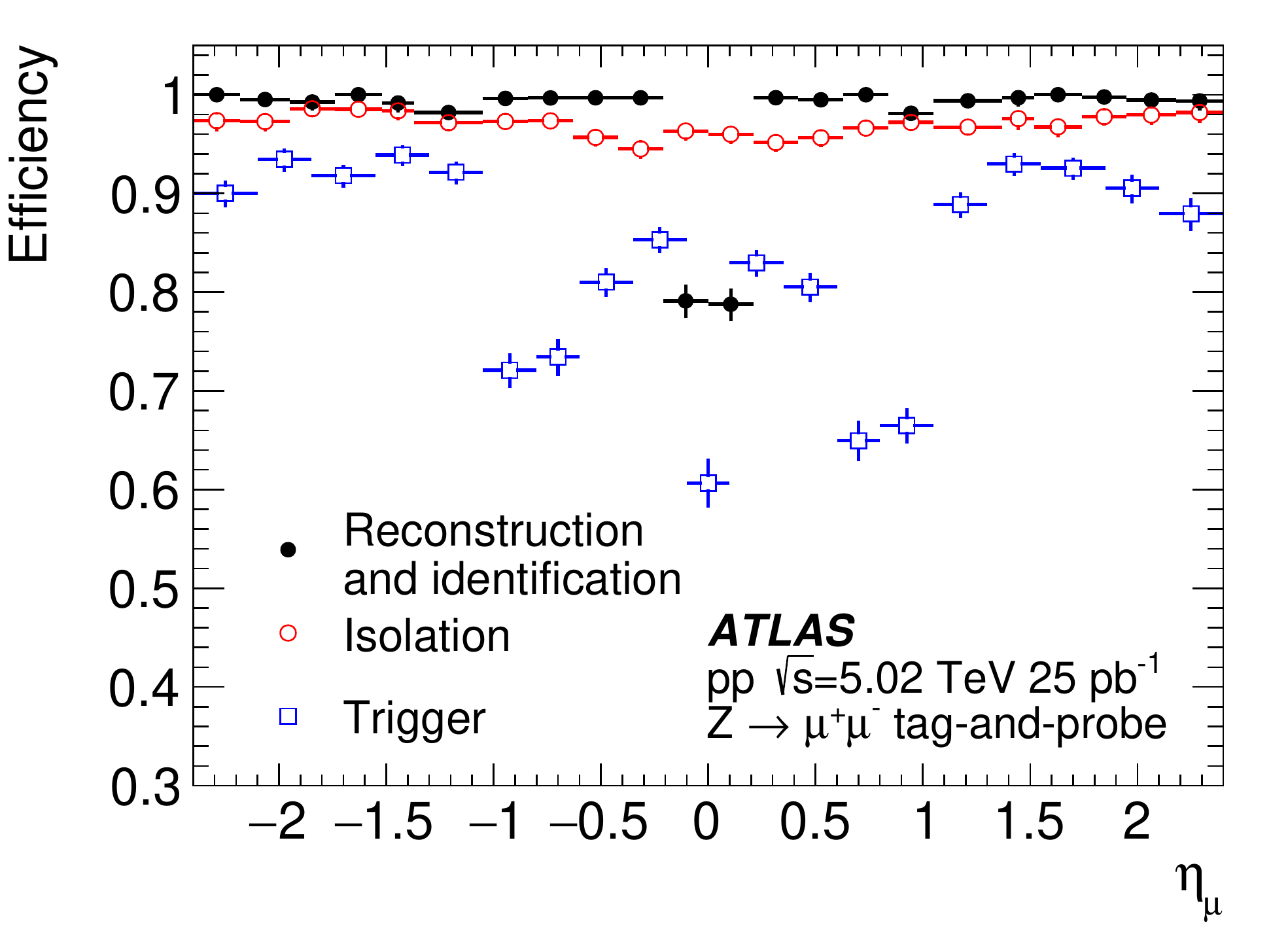}}
  \caption{Efficiencies of reconstruction, identification, isolation and trigger requirements as a function of lepton pseudorapidity for (a)~electrons and (b)~muons measured using the tag-and-probe method. The efficiency of each selection is defined with respect to leptons selected in the previous step. \label{fig:lept-perf}}
\end{figure}

Figure~\ref{fig:lept-perf} summarises the reconstruction, identification, isolation and trigger efficiencies for electron and muon candidates obtained from the tag-and-probe method.

\begin{figure}[!ht]
  \centering
  \subfloat[]{\includegraphics[width=0.49\textwidth]{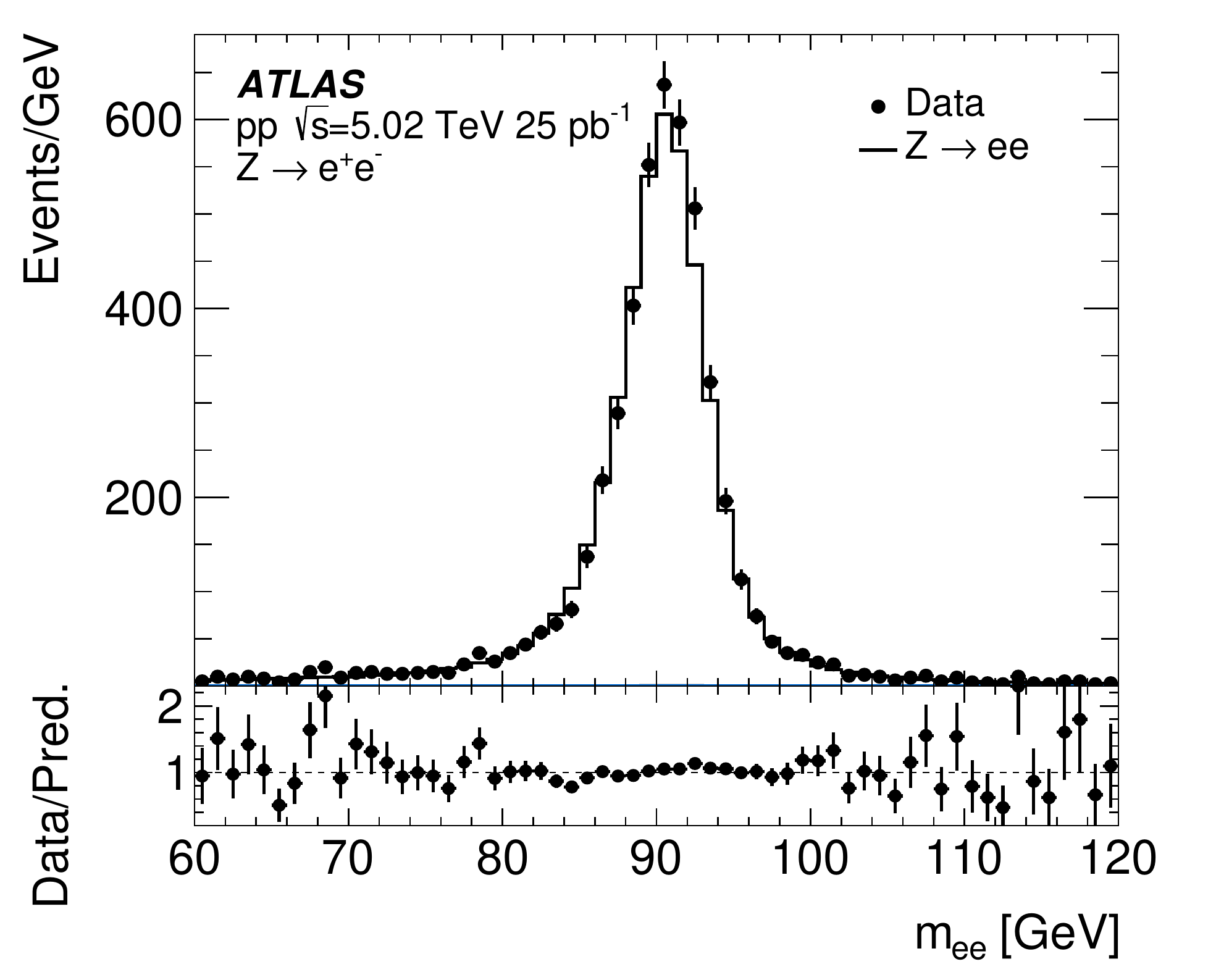}}
  \subfloat[]{\includegraphics[width=0.49\textwidth]{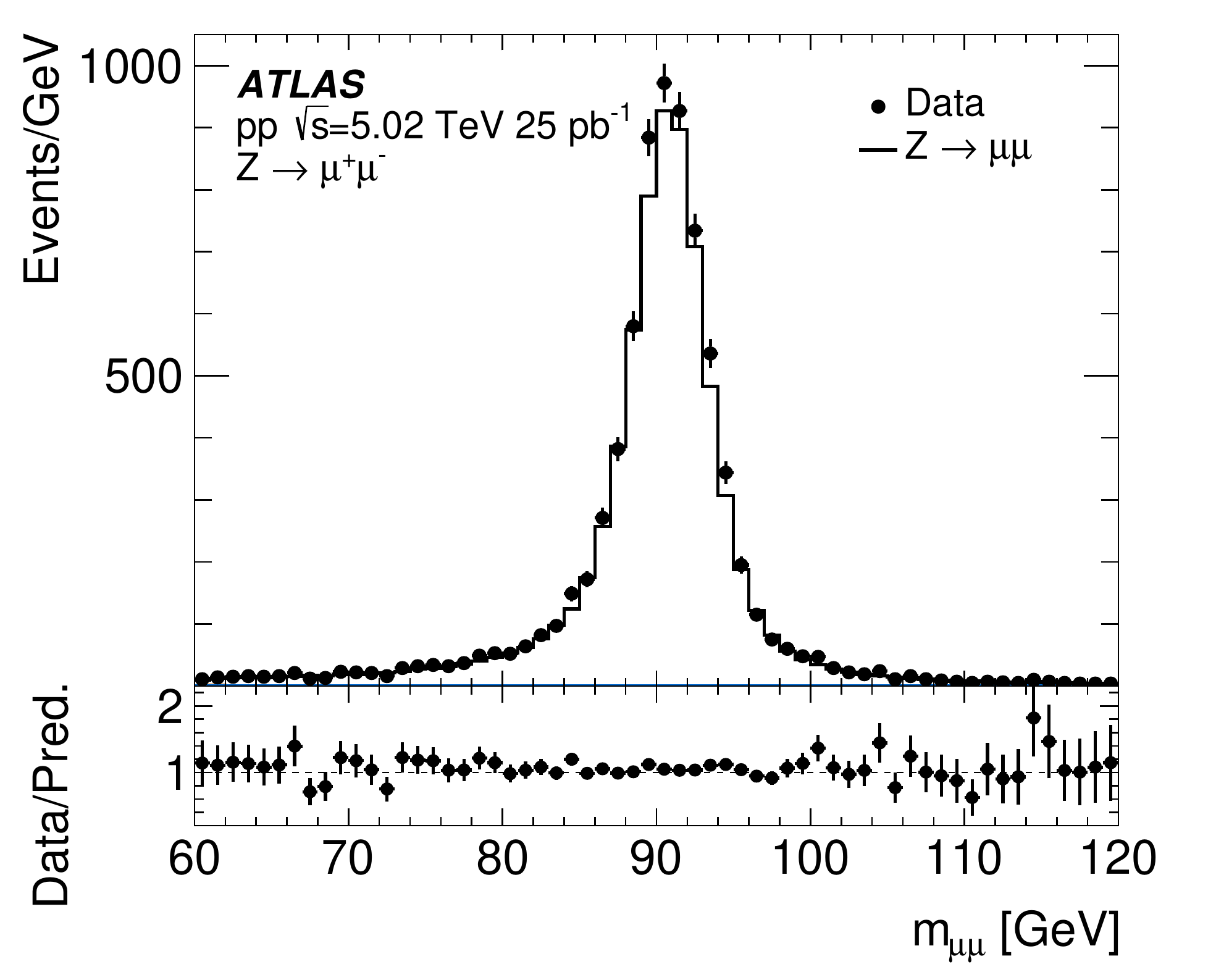}} 
  \caption{Detector-level invariant mass distribution of (a)~dielectron and (b)~dimuon pairs from $Z$ boson decays together with EW background contributions. Background contributions are too small to be visible on a linear scale. Only the statistical uncertainties of the data are shown. \label{fig:zmass}}
\end{figure}

Figure~\ref{fig:zmass} shows the invariant mass distribution of the dilepton system for electron and muon candidates from \zll\ boson decays after applying scale factors to the MC simulation. The data points are compared with simulation including $Z$ boson signal and background components. The electron candidates in the data, shown on the left panel, are calibrated using calorimeter settings and calibration correction factors optimised for low-pile-up conditions. Good agreement between the data and the simulation is found for both channels.

\subsection{Recoil calibration}

In events with $W$ or $Z$ boson production, the hadronic recoil gives a measure of the boson transverse momentum. The calibration of the recoil is performed
using dilepton events from decays of $Z$ bosons produced in $pp$ collisions at $\sqrt{s} = 5.02$~\TeV, as information about the $Z$ boson transverse momentum 
can be obtained with high precision from the measurements of lepton momenta and compared with the measurement from hadronic recoil.
The recoil resolution is studied using $u_{\perp}$, the projection of \utvec\ onto the axis -- in the transverse plane -- perpendicular to the $Z$ boson \ptvec. The resolution is given by the standard deviation of the $u_{\perp}$ distribution, $\sigma_{u_{\perp}}$. The transverse momentum scale response of the recoil
can be studied using the bias defined as $u_{\parallel} + \pt^Z$, where $u_{\parallel}$ is the projection of \utvec\ onto the axis defined by $\ptvec^Z$, and is quantified via the average of the bias distribution.
Differences between the responses in data and simulation are less than ${\sim} 2$~\GeV, while up to ${\sim} 20\%$ differences 
in the resolution are observed.

\begin{figure}[!ht]
  \centering
  \subfloat[]{\includegraphics[width=0.49\textwidth]{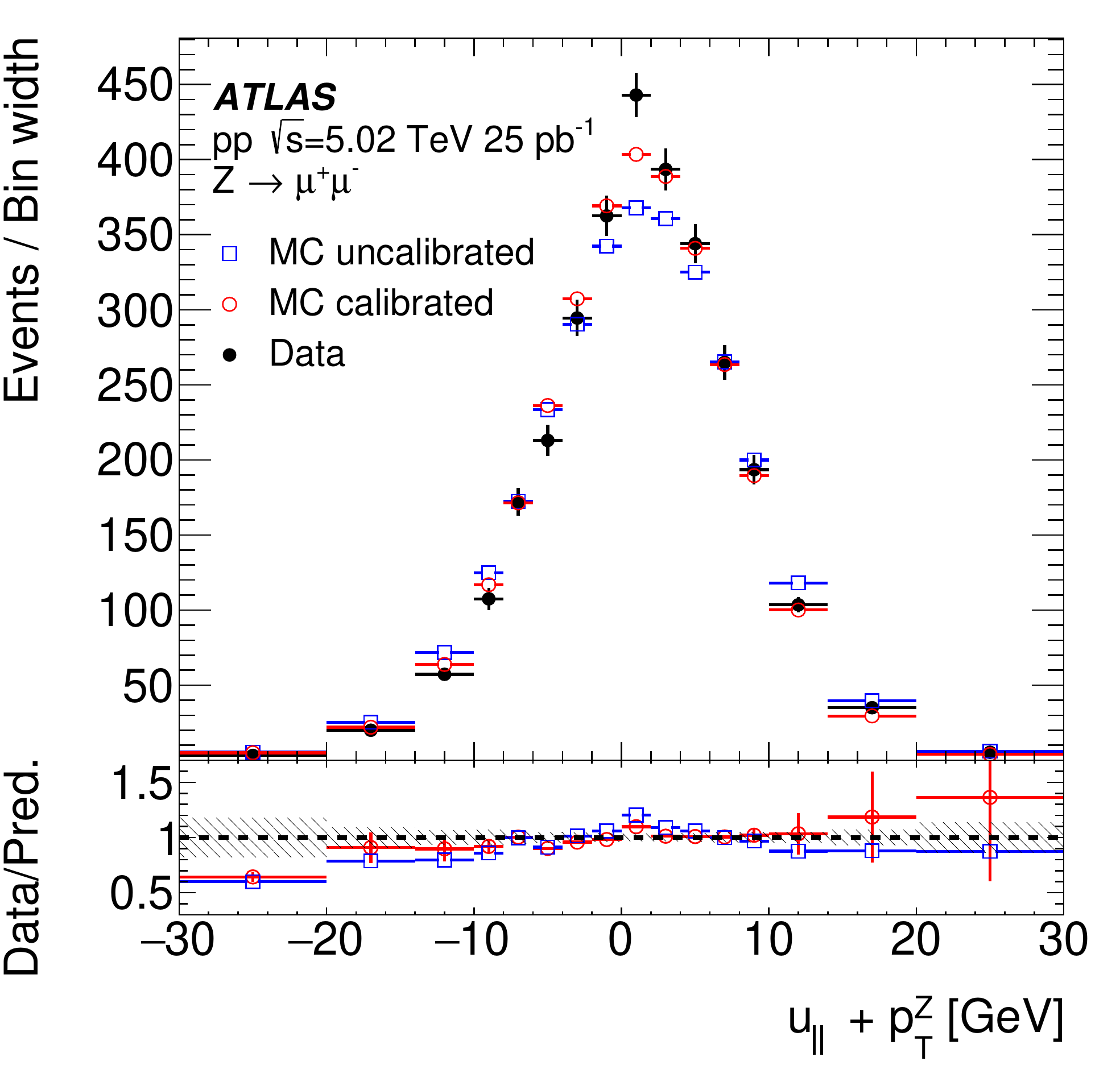}}
  \subfloat[]{\includegraphics[width=0.49\textwidth]{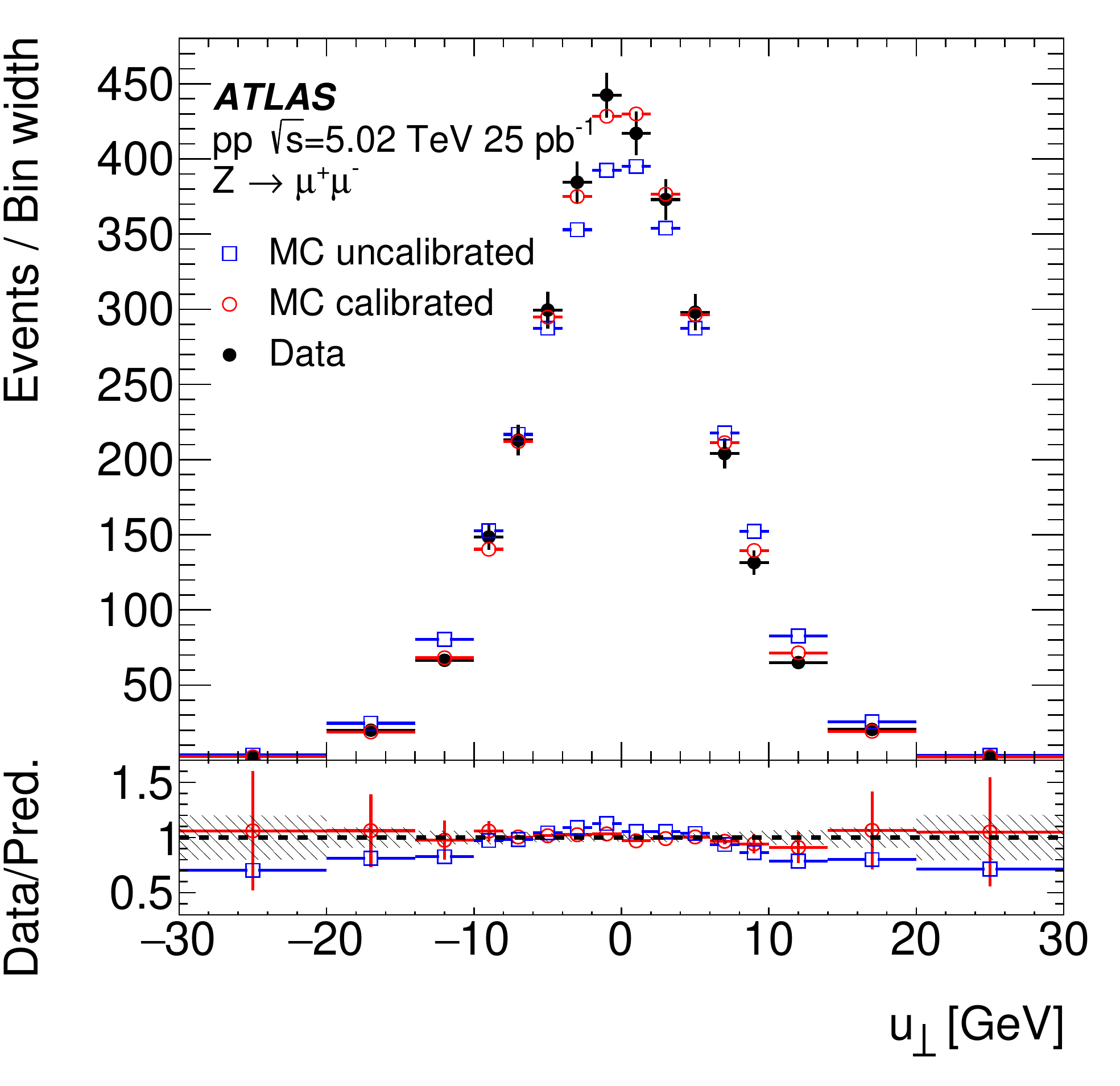}}
  \caption{Distributions of (a)~$u_{\parallel} + \pt^Z$ and (b)~$u_{\perp}$ in data and \zmm\ MC simulation before~(squares) and after~(circles) recoil calibration. The shaded band in the ratio panels represents the statistical uncertainty of the data sample, while the error bars represent the systematic uncertainty associated with the calibration procedure.}
  \label{fig:hr-calib}
\end{figure}

Following the procedure described in Ref.~\cite{Aaboud:2017svj}, \textit{in situ} corrections to the resolution and the scale of \ut\ are obtained in $Z$ events and are applied to the $W$ boson event candidates, as a function of $\pt^W$.
The corrections applied to the simulation are obtained as a function of $\pt^Z$:
\begin{eqnarray}
\label{eq:reccorr1} u^{W,\textrm{corr}}_{\parallel} &=& \langle u^{Z}_{\parallel} + \pt^Z \rangle^{\textrm{data}} - \langle u^{Z}_{\parallel} + \pt^Z \rangle^{\textrm{MC}} + \langle u^{Z,\textrm{data}}_{\parallel} \rangle +  \left( u^{W,\textrm{MC}}_{\parallel} - \langle u^{Z,\textrm{data}}_{\parallel} \rangle \right) \cdot \frac{\sigma_{u_{\perp}}^{\textrm{data}}}{\sigma_{u_{\perp}}^{\textrm{MC}}};\\
\label{eq:reccorr2} u^{W,\textrm{corr}}_{\perp} &=& u^{W,\textrm{MC}}_{\perp} \cdot \frac{\sigma_{u_{\perp}}^{\textrm{data}}}{\sigma_{u_{\perp}}^{\textrm{MC}}}.
\end{eqnarray}
Equation~\eqref{eq:reccorr1} describes corrections applied to the recoil response in simulation. It includes a shift which brings the average value of $u_{\parallel}$ in the simulation closer to the one in data, taking into account differences in the bias. In addition, it corrects the response distribution for resolution differences (last term in the equation). The resolution correction is directly described by Eq.~\eqref{eq:reccorr2} where it is applied to the $u_{\perp}$ distribution in the simulation.
The impact of the calibration on the scale and resolution in events where a $Z$ boson decays to a dimuon pair is shown in Figure~\ref{fig:hr-calib}. The distributions are shown for the simulation before and after applying the corrections and for data. 
Agreement of the distributions from simulation with data distributions is improved after applying the calibration, and residual differences are covered by the systematic uncertainties described in Section~\ref{sec:uncertainties}.

\section{Background determination \label{sec:bkg}}

\subsection{$W$ channels}

The reported cross-section measurements correspond to inclusive Drell--Yan production of single vector bosons which decay leptonically. Background processes that contribute to the \Wpm~boson production measurement are EW processes producing \wtnu, \zll, \ztt\ decays, EW diboson ($WW$, $WZ$, $ZZ$) production, as well as top-quark production and multi-jet processes. The multi-jet background includes various processes such as semileptonic decays of heavy-flavour hadrons or in-flight decays of kaons and pions for the muon channel, as well as photon conversions or misidentified hadrons for the electron channel. The background contributions from EW and top-quark production are evaluated using simulated event samples, while the multi-jet contribution is estimated with a data-driven method similar to the one described in Ref.~\cite{Aad:2016naf}.

Although multi-jet background events are well rejected by the lepton isolation requirements, their contribution to the signal region is still sizeable because of the very large production cross sections for multi-jet processes. This contribution is estimated from template fits to data in kinematic distributions: lepton \pt, \met\ and \mt. The fits are performed in a phase-space region defined by the full event selection with a looser lepton \pt\ requirement of $\pt > 20$~\GeV\ and with the requirements on \met\ and \mt\ 
removed. An additional requirement on the transverse component of the hadronic recoil, $\ut < 30$~\GeV,
is placed to ensure better agreement of the event kinematics between the fit region and the signal region.

\begin{figure}[!ht]
  \centering
  \subfloat[]{\includegraphics[width=0.49\textwidth]{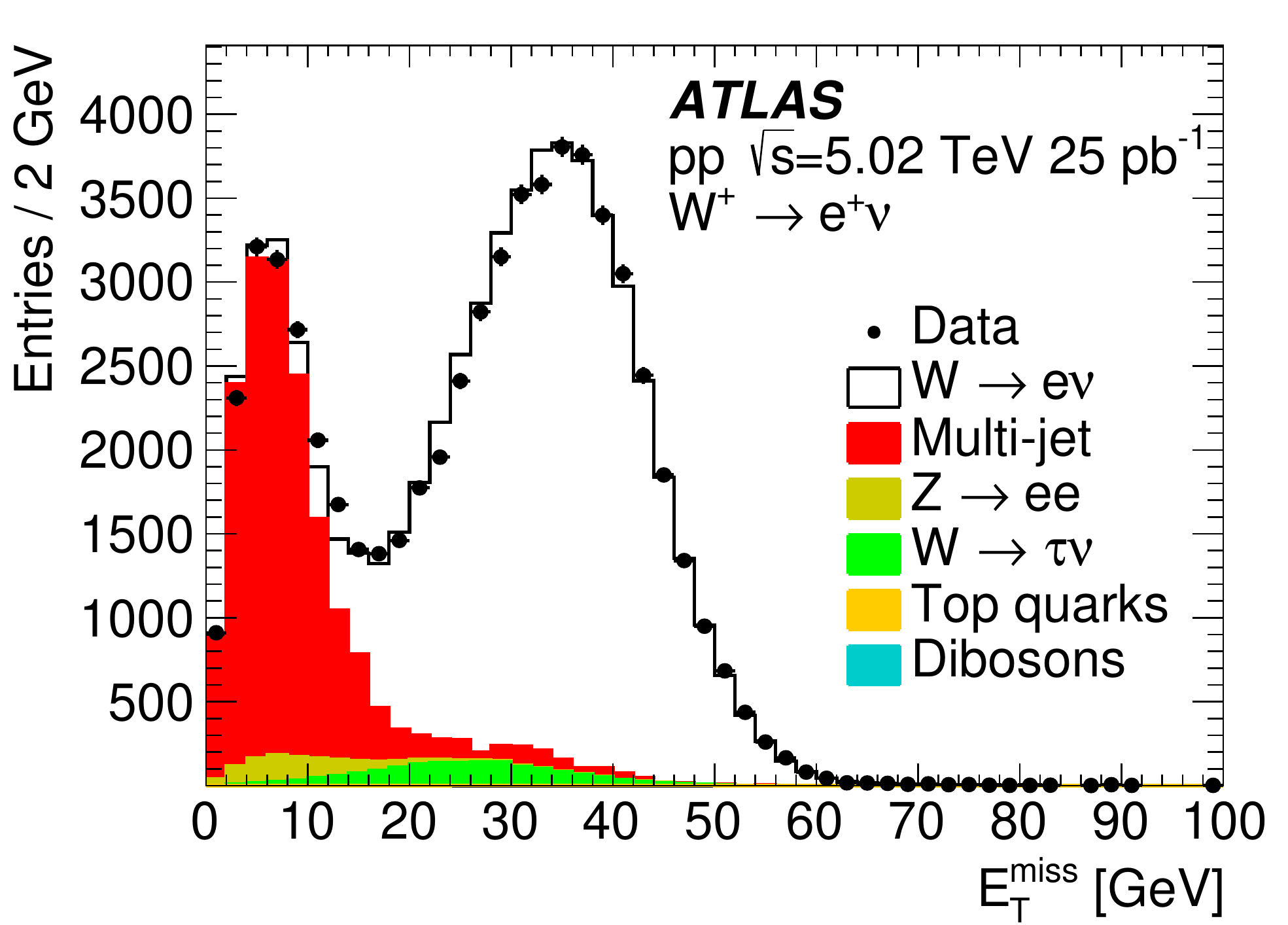}}
  \subfloat[]{\includegraphics[width=0.49\textwidth]{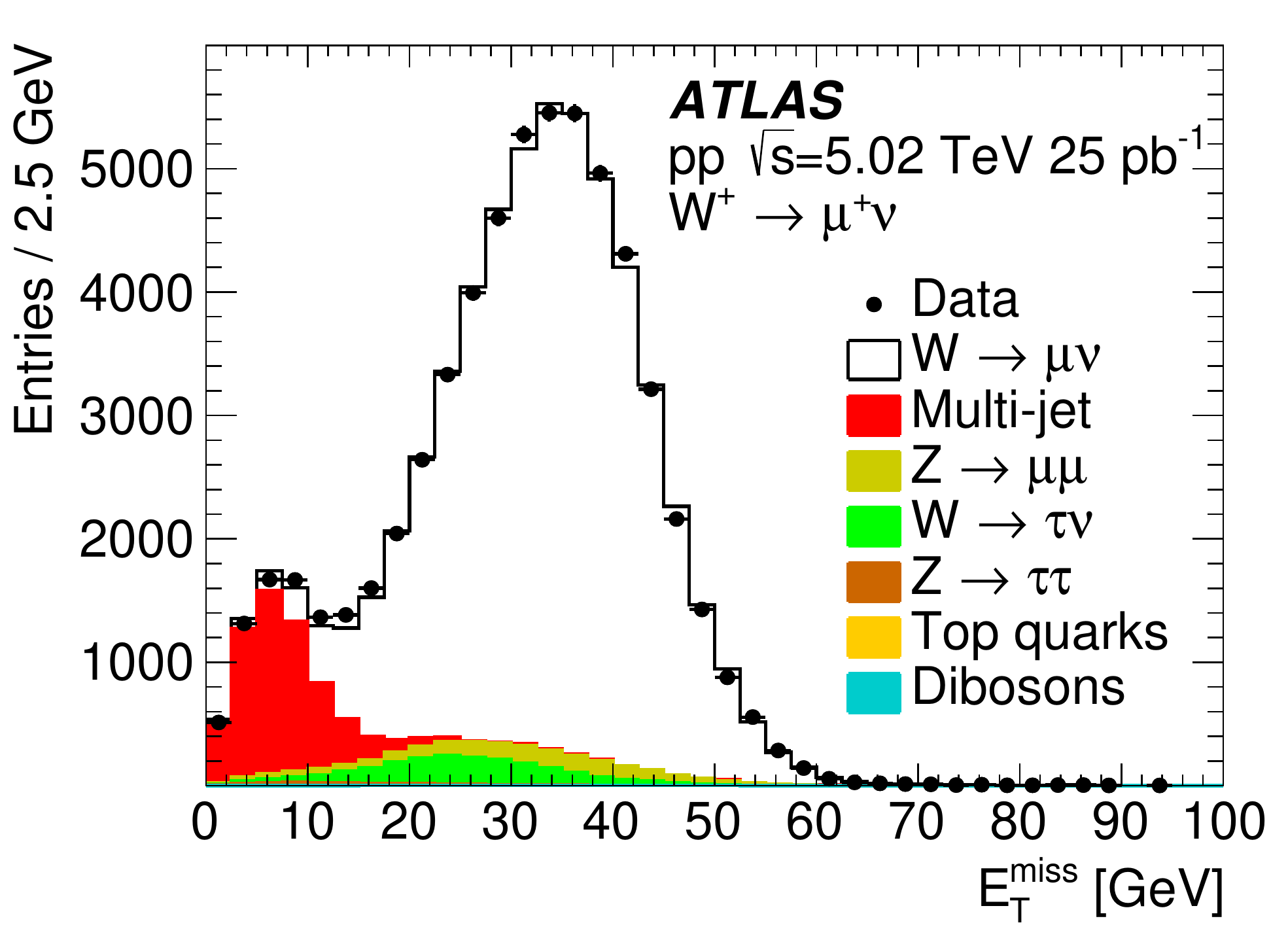}}
  \caption{Distributions of \met\ used to extract multi-jet yields in the (a)~electron and (b)~muon channels after performing the template fits. Only the statistical uncertainties of the data are shown. \label{fig:bkg-fit}}
\end{figure}

Template distributions for signal, EW and top-quark background processes are constructed by applying the fit-region selection to samples of simulated events. Templates enriched in contributions from multi-jet processes are built using events in data with non-isolated leptons selected by inverting the isolation requirement described in Section~\ref{sec:selection}. 
The normalisation factors of template distributions for signal, EW and top-quark backgrounds, as well as the multi-jet background, are extracted from a fit to the data. The fits are repeated with multi-jet background templates 
constructed from different intervals in a track-based (muon channel) or calorimeter-based (electron channel) isolation variable. Finally, a linear extrapolation to the signal region is performed as a function of the selected isolation variable, accounting also for the difference in kinematic selections between the fit region and the signal region.
Examples of post-fit template \met\ distributions, which are used to extract multi-jet yields in the electron and muon channels, are presented in Figure~\ref{fig:bkg-fit}.

Following this procedure, multi-jet background processes are estimated to contribute around 0.9\% of the \wpenu\ sample and 1.4\% of the \wmenu\ sample, while in the muon channel they represent around 0.1\% of the \wpmunu\ sample and 0.2\% of the \wmmunu\ sample.

The largest background contributions to the decay modes studied come from the production of single EW bosons decaying via 
other decay channels. The \zee\ background represents 0.1\% of the \wpenu\ sample and 0.2\% of the \wmenu\ sample, while the \zmm\ background amounts to~2.8\% and~3.8\% in the \wpmunu\ and \wmmunu\ samples, respectively. The \wtnu\ background contributes around~1.8\% to the samples selected in both channels and the \ztt\ background contributes approximately~0.1\%. Contributions from top-quark production ($t\bar{t}$ and single top quarks) are estimated to be at the level of~0.1--0.2\% in both channels. Similarly, diboson processes represent approximately~0.1\% of the selected event samples.

\begin{figure}[!ht]
  \centering
  \subfloat[]{\includegraphics[width=0.49\textwidth]{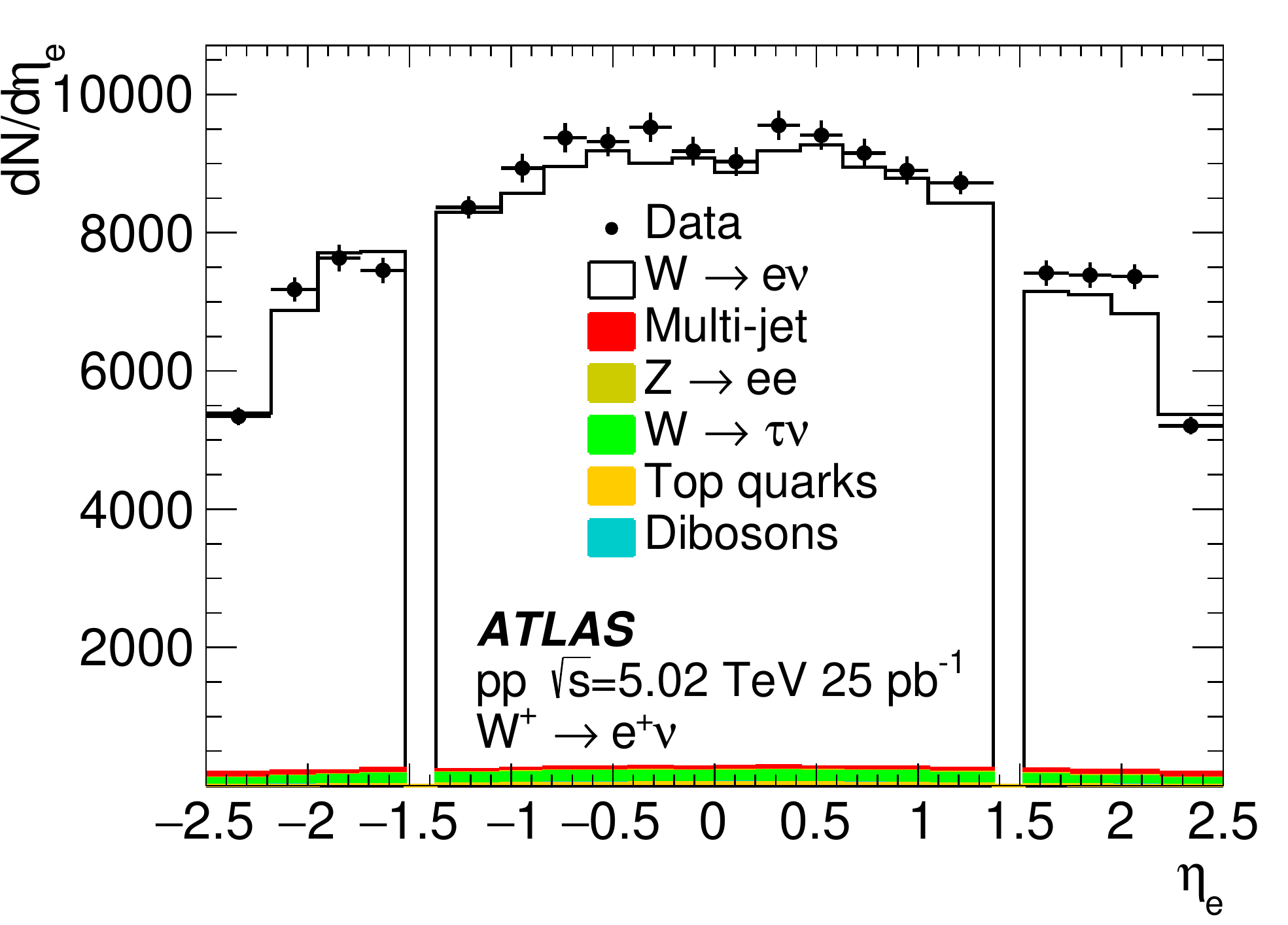}}
  \subfloat[]{\includegraphics[width=0.49\textwidth]{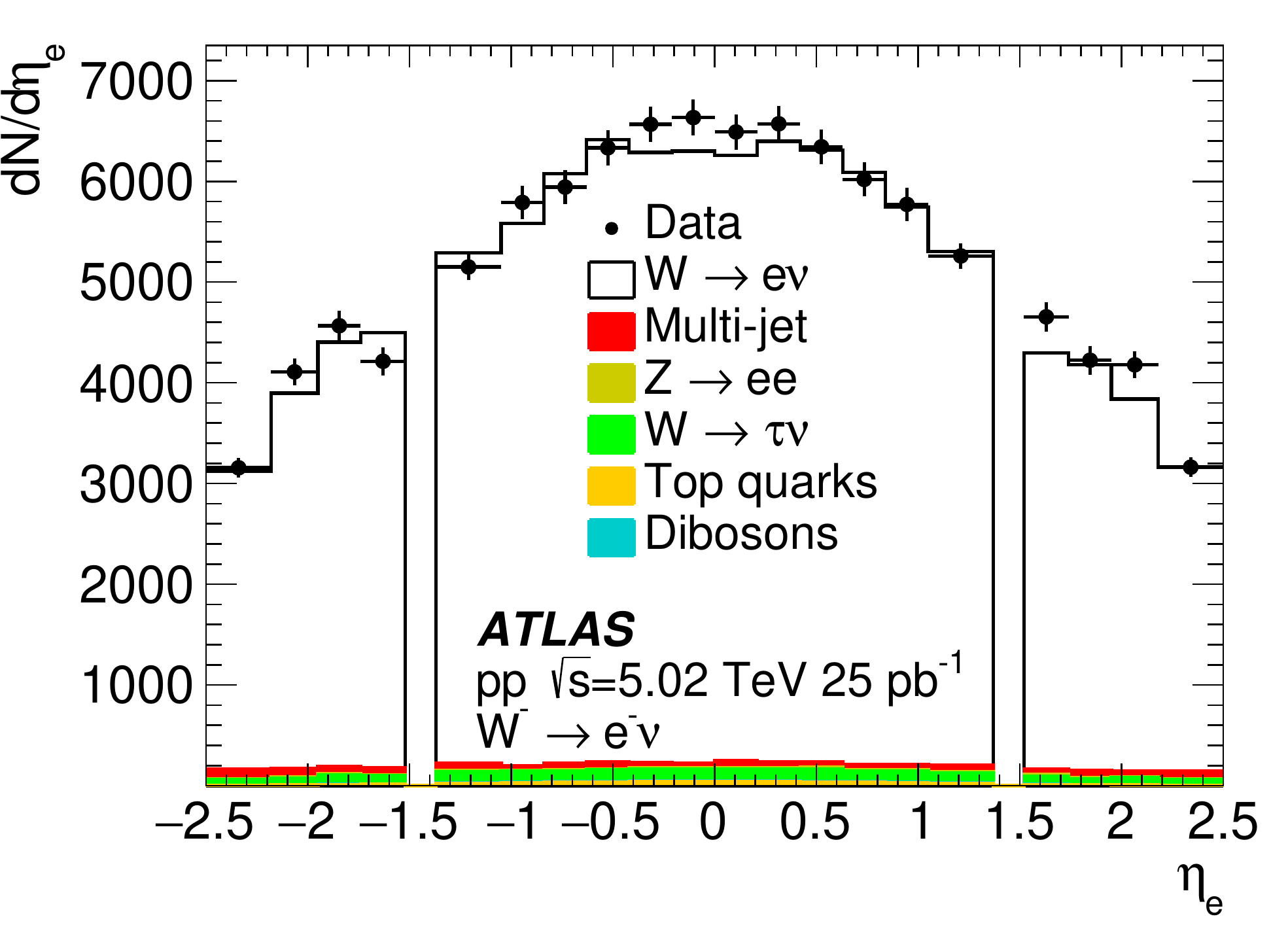}}
  \caption{Distribution of detector-level lepton pseudorapidity for (a)~\wpenu\ and (b)~\wmenu. Only the statistical uncertainties of the data are shown. \label{fig:control-we}}
\end{figure}

\begin{figure}[!ht]
  \centering
  \subfloat[]{\includegraphics[width=0.49\textwidth]{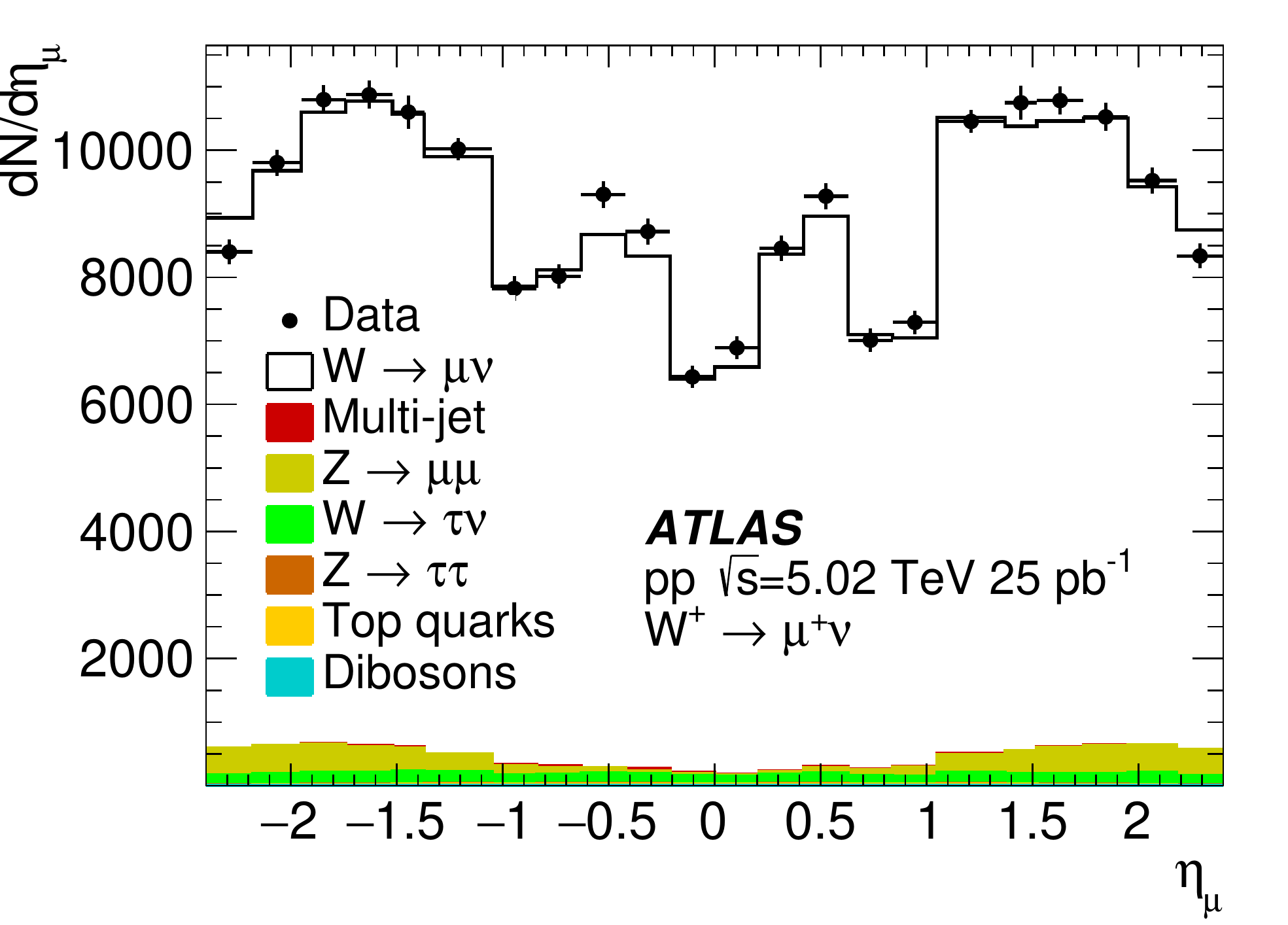}}
  \subfloat[]{\includegraphics[width=0.49\textwidth]{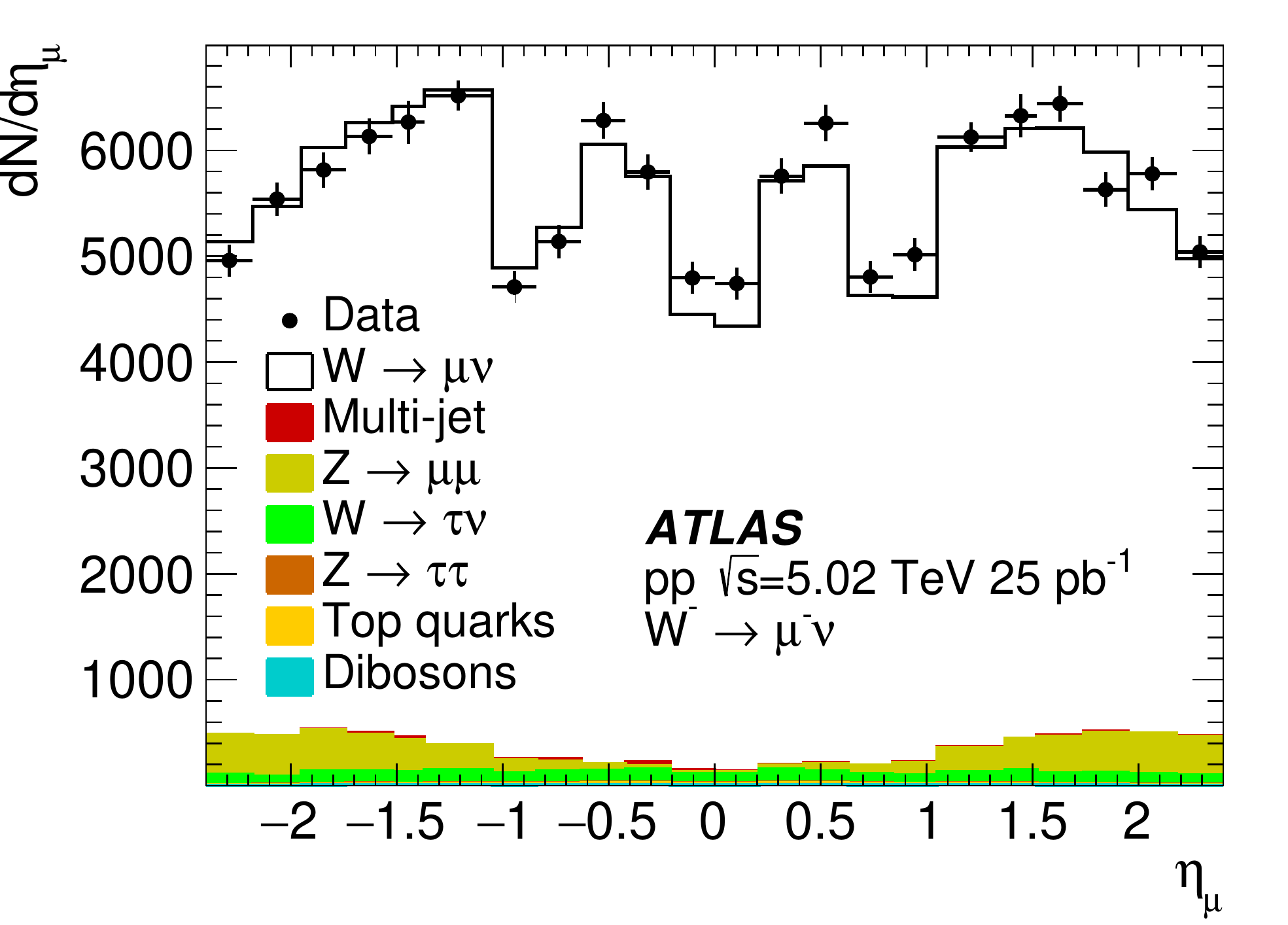}}
  \caption{Distribution of detector-level lepton pseudorapidity for (a)~\wpmunu\ and (b)~\wmmunu. Only the statistical uncertainties of the data are shown. \label{fig:control-wmu}}
\end{figure}

Figures~\ref{fig:control-we} and~\ref{fig:control-wmu} show detector-level lepton pseudorapidity distributions for positive 
and negative electron and muon candidates from $W$ boson decays. Good agreement is found between the data and the sum of signal and background contributions.

\subsection{$Z$ channels}

Background contributions to the $Z$ boson sample are expected from \ztt, diboson and $W$ boson decay processes, top-quark pair production, and the multi-jet background. The EW and top-quark contributions are evaluated from dedicated simulation samples, whereas the upper limit on the amount of the multi-jet background is estimated.

Diboson background contributes 0.08\% in the muon channel and 0.14\% in the electron channel. The \ztt\ background is found to be at the level of 0.07\% in both decay channels. The top-quark background is at the level of 0.06\% in the electron channel and 0.08\% in the muon channel. The $W$ boson background is found to be below 0.01\% in both channels.

The contribution of the multi-jet background in the muon channel is estimated from samples that simulate $b\bar{b}$ and $c\bar{c}$ production. The study yields an estimate at the level of <0.01\%. A previous ATLAS measurement at $\sqrt{s}=7$~\TeV~\cite{Aaboud:2016btc} estimated the multi-jet contribution at the level of 0.02--0.15\% for the electron channel and 0.09\% for the muon channel. As it is expected that this contribution increases with pile-up and since that measurement was done with higher pile-up than the current analysis, the multi-jet background is considered to be negligible in this analysis.

\begin{figure}[!ht]
  \centering
  \subfloat[]{\includegraphics[width=0.49\textwidth]{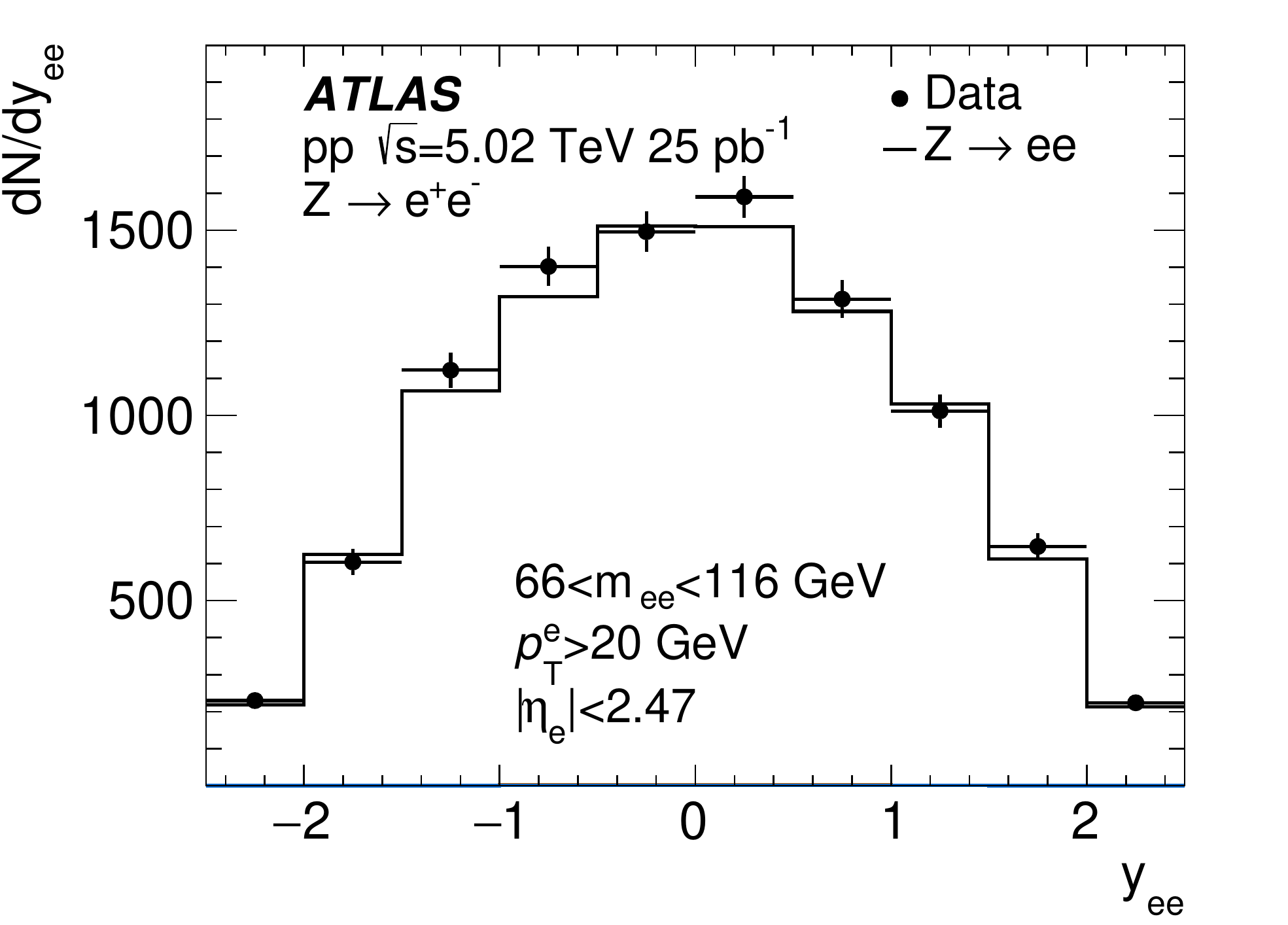}}
  \subfloat[]{\includegraphics[width=0.49\textwidth]{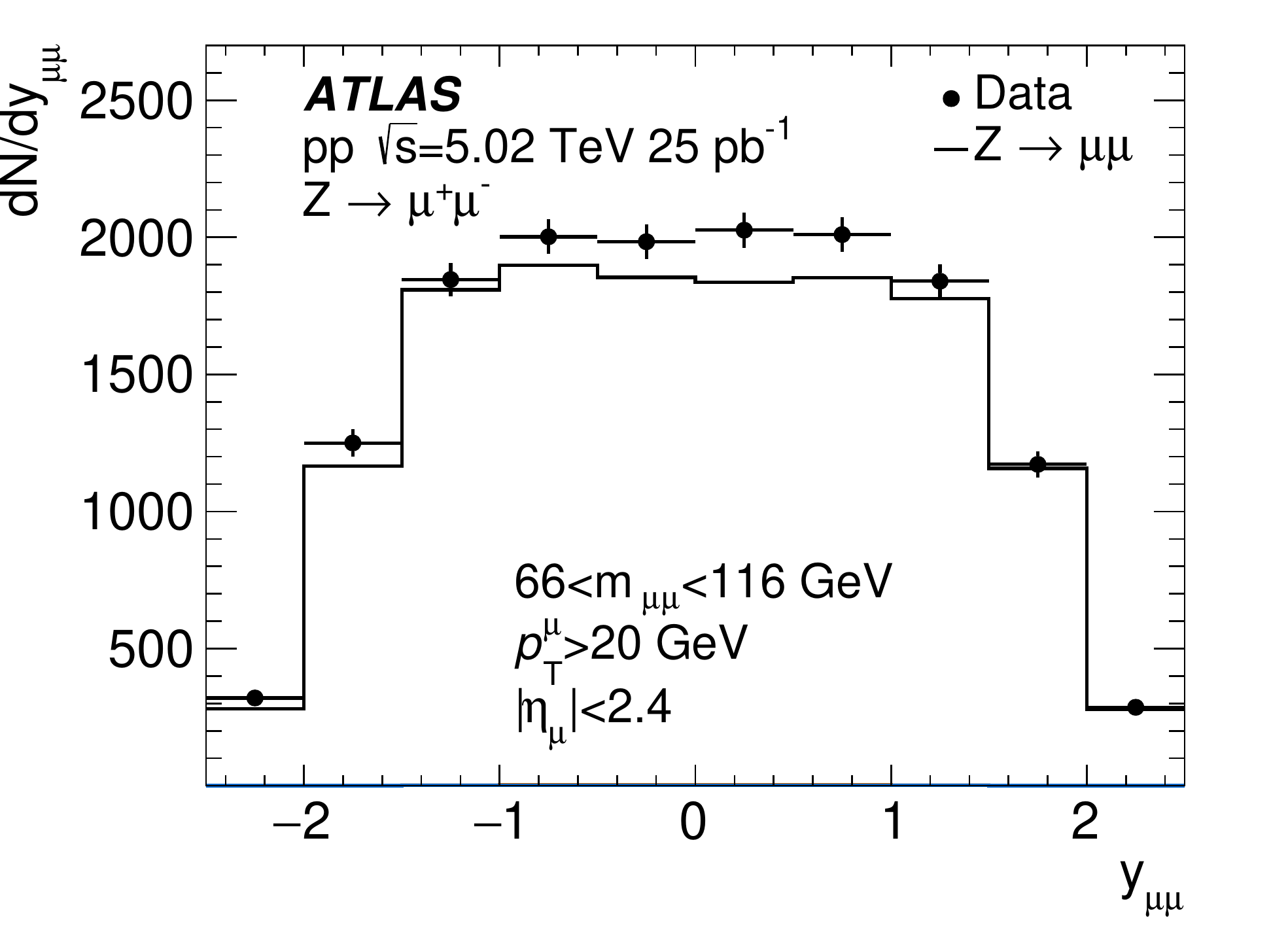}}
  \caption{Detector-level lepton-pair rapidity distributions in the (a)~electron and (b)~muon channels. Background contributions are negligible using a linear scale. Only the statistical uncertainties of the data are shown. \label{fig:control-z}}
\end{figure}

Figure~\ref{fig:control-z} shows detector-level dilepton rapidity distributions for electron and muon candidates from $Z$ boson decays.
Good agreement is found between the data and the sum of signal and background contributions.

Table~\ref{tab:bkg} summarises background contributions to the $W^+$, $W^-$ and $Z$ boson candidate samples.  
\begin{table}
  \centering
  \caption{Background contributions as a percentage of the total for the $W^+$, $W^-$ and $Z$ candidate samples in the electron~(muon) channels.\label{tab:bkg}}
  \begin{tabular}{l|ccc}
    \hline
    \hline
    Background & \wpenu~(\wpmunu) &\wmenu~(\wmmunu) & \zee (\zmm)\\
     & [\%] & [\%] & [\%]\\
    \hline
    \zll, $\ell=e,\mu$ &0.1 (2.8) &0.2 (3.8) & --\\
    \wlnu, $\ell=e,\mu$ & -- & -- & $<$0.01 ($<$0.01)\\
    \wtaunu & 1.8 (1.8) & 1.8 (1.8) & $<$0.01 ($<$0.01)\\
    \ztautau & 0.1 (0.1) & 0.1 (0.1) & \;\;0.07 \;\;(0.07)\\
    Multi-jet & 0.9 (0.1) & 1.4 (0.2) & $<$0.01 ($<$0.01)\\
    Top quark & 0.1--0.2 (0.1--0.2) & 0.1--0.2 (0.1--0.2) & \;\;0.06 \;\;(0.08)\\
    Diboson & 0.1 (0.1) & 0.1 (0.1) & \;\;0.14 \;\;(0.08)\\
    \hline\hline
  \end{tabular}
\end{table}

\section{Measurement procedure \label{sec:procedure}}

The integrated and differential $W$ and $Z$ boson production cross sections are measured within a fiducial phase space defined as follows:
\begin{itemize}
\item for $W$ production : $\ptl>25$~\GeV, $\ptnu>25$~\GeV, $|\etal|<2.5$, $\mt>40$~\GeV
\item for $Z$ production : $\ptl>20$~\GeV, $|\etal|<2.5$, $66<\mll<116$~\GeV.
\end{itemize}
Integrated fiducial cross sections in the electron and muon channels are calculated using:
\begin{equation}
  \sigma^\mathrm{fid}_{\wlnu[\zll]} = \frac{N_{W[Z]} - B_{W[Z]}}{\CWZ \cdot L_\mathrm{int}} \,,
\label{eq:WZxsecfidemu} 
\end{equation}
where $N_{W[Z]}$ and $B_{W[Z]}$ are the number of selected events in data and the expected number of background events, respectively.
The integrated luminosity of the sample is $L_\mathrm{int}=\lumiprecise\pm\dlumiabs\,\ipb$, determined with the method described in Ref.~\cite{Aaboud:2016hhf}.
A correction for the event detection efficiency is applied with the factor \CWZ\ , which is obtained from the signal simulation described in Sec.~\ref{sec:samples} as:
\begin{equation}
\nonumber
  \CWZ = \frac{N^\mathrm{MC, sel}_{W[Z]}}{N^\mathrm{MC, fid}_{W[Z]}}\,.
\label{eq:CWZ}
\end{equation}
Here, $N^\mathrm{MC, sel}_{W[Z]}$ is the number of events which pass the signal selection at the detector level, corrected for the observed differences between data and simulation such as in 
reconstruction, identification, isolation, and trigger efficiencies. The denominator $N^\mathrm{MC, fid}_{W[Z]}$ is computed applying the fiducial requirements to the generator-level leptons originating from $W$ and $Z$ boson decays. The measurement is corrected for QED final-state radiation effects by applying these requirements to the lepton momenta before photon radiation.
The \CWZ\ factors also account for the difference in acceptance between detector-level requirements on lepton~$|\eta|$ and the fiducial selection of $|\etal|<2.5$.

The procedure described above is extended to the measurement of differential cross sections as a function of the decay lepton pseudorapidity in $W$ boson production, and as a function of the lepton-pair rapidity in $Z$ boson production. The dependence of cross sections on these kinematic variables is particularly sensitive to the choice of PDFs. For the measurement of differential cross sections, the formula given in Eq.~\eqref{eq:WZxsecfidemu} is adjusted so that the cross sections are divided by the width of the corresponding interval in absolute pseudorapidity or rapidity. For $W$ production, following Ref.~\cite{Aaboud:2016btc}, the lepton $|\eta|$ boundaries are defined as:
\begin{itemize}
\item 0 -- 0.21 -- 0.42 -- 0.63 -- 0.84 -- 1.05 -- 1.37 -- 1.52 -- 1.74 -- 1.95 -- 2.18 -- 2.50;
\end{itemize}
for $Z$ boson production, the lepton-pair $|\yll|$ boundaries are defined as:
\begin{itemize}
\item 0 -- 0.5 -- 1.0 -- 1.5 -- 2.0 -- 2.5.
\end{itemize}
For the measurement of these cross sections, the \CWZ\ factors are computed separately for each lepton $|\eta|$ or $|\yll|$ interval by applying the corresponding requirements on the reconstructed lepton kinematics in the numerator, and on the generator-level kinematics in the denominator. Migrations between rapidity intervals are negligible due to the very good angular resolution with which charged-particle tracks associated with leptons are reconstructed, and the good lepton momentum and energy resolutions. The \CWZ\ factors for the measurements of integrated and differential cross sections are summarised in Table~\ref{tab:CWZ}.

The uncertainty associated with the \CWZ\ correction is dominated by experimental systematic uncertainties, described in Section~\ref{sec:uncertainties}. For the differential \CW\ factors, the relative size of statistical and systematic uncertainties added in quadrature varies in the range 1.1--2.5\% (1.7--3\%), while the uncertainties in differential \CZ\ factors are in the range 1.6--3.5\% (0.9--1.2\%) in the electron (muon) channel.

Uncertainties in \CWZ\ of theoretical origin comprise uncertainties induced by the PDFs, by the description of the $W$ and $Z$ boson transverse momentum distributions, by the implementation of the NLO
QCD matrix element and its matching to the parton shower, and by the modelling of the parton shower, hadronisation and underlying event. These uncertainties are discussed
in Ref.~\cite{Aaboud:2016btc}, where they are evaluated to be smaller than 0.2\% and thus are negligible at the present level of precision.
The size of acceptance corrections included in the integrated correction factors is 7\% (3\%) for the $W$ boson measurements and 14\% (5\%) for the $Z$ boson measurements in the electron (muon) channel. In the case of differential $W$ boson measurements, only the \CW\ factor in the interval $2.18 < |\etal| <2.5$ includes an acceptance correction of 9\% for \wenu\ processes and 40\% for \wmunu\ processes. On the other hand, all differential \CZ\ factors include an acceptance correction which varies from 6\% to 28\% for the \zee\ channel and up to 53\% for the \zmm\ channel.

\begin{table}
\centering
\caption{Correction factors \CWZ\ used to calculate integrated and differential $W$ and $Z$ boson production cross sections. The integrated \CWZ\ factors are shown with the sum in quadrature of statistical and systematic uncertainties. For the differential \CWZ\ factors, the spread of values across lepton $|\eta|$ or $|\yll|$ intervals is shown, while their uncertainties are described in the text. \label{tab:CWZ}}
\begin{tabular}{l|ccc}
\hline
\hline
Channel			& \CW\ (\wplnu)	& \CW\ (\wmlnu)	& \CZ \\
\hline
\multicolumn{4}{l}{Integrated cross-section measurements} \\
\hline
Electron channel	& 0.657 $\pm$ 0.006	& 0.667 $\pm$ 0.005	& 0.522 $\pm$ 0.007 \\
Muon channel		& 0.723 $\pm$ 0.011	& 0.720 $\pm$ 0.010	& 0.780 $\pm$ 0.007 \\
\hline
\multicolumn{4}{l}{Differential cross-section measurements} \\
\hline
Electron channel	& \multicolumn{2}{c}{0.55--0.80}	& 0.52--0.62 \\
Muon channel		& \multicolumn{2}{c}{0.55--0.85}	& 0.60--0.82 \\
\hline\hline
\end{tabular}
\end{table}

\section{Measurement uncertainties \label{sec:uncertainties}}
\subsection{Lepton calibration and efficiency corrections}
Uncertainties in the determination of lepton trigger, reconstruction, identification and isolation efficiency scale factors affect the measurements through the correction factors \CWZ.

The uncertainties of the electron efficiency measurements are divided into contributions correlated across electron $\eta$ and \pt\ intervals and uncorrelated ones, and are propagated to the cross-section measurements accordingly.
For the \wenu\ channels the efficiency determination contributes a systematic uncertainty of 0.8\% to the fiducial cross-section measurements, while for the \zee\ channel this
contribution is 1.3\%. Systematic effects related to the electron \pt\ scale and resolution are subdominant, yielding an uncertainty at the level of 0.3\% for the \wenu\ channels and less than 0.2\%
for the \zee\ channel. Uncertainties in the modelling of the electron charge identification are at the level of 0.1\%, and neglected for the cross section measurements. Their impact on the asymmetry measurements is however sizeable and included in the final results.

In the muon channels, the statistical components of the scale factor uncertainties are propagated to the measurements via MC pseudo-experiments, while systematic components are propagated as a single variation fully correlated across all muon $|\eta|$ intervals. The single largest contribution to the systematic uncertainty of fiducial cross-section measurements in the \wmunu\ channels is 1.4\%\ and comes from the determination of the muon trigger efficiency. 
For measurements in the \zmm\ channel the largest systematic uncertainty is contributed by the muon isolation efficiency measurement and amounts to 0.7\%. Uncertainties coming from the muon \pt\ scale and resolution are below 0.2\% for both \wmunu\ channels and the \zmm\ channel.

\subsection{Hadronic recoil corrections}
The uncertainty assigned to the hadronic recoil calibration is conservatively defined from the full size of the corrections, which are derived using events with \Zboson\ boson production. In these events, the impact of the correction on the $u_{\perp}$ and $u_{\parallel}+\pt^Z$ distributions 
varies between a few percent and ${\sim}20\%$ in the range $[-15, +15]$~\GeV, 
which dominates the reported cross-section measurements. After applying this correction to events with \Wplus\ and \Wminus\ production, the resulting uncertainties on the cross-section measurements are at the level of 0.5\% for both the muon and electron channels.

\subsection{Background evaluation}

Uncertainties in the evaluation of EW and top-quark backgrounds in the \wenu\ and \wmunu\ channels are estimated by varying the respective normalisation 
cross sections. For single-boson production, the size of the cross-section variations is obtained from higher-order QCD calculations, while for diboson and 
top-quark processes the uncertainty in the cross sections is conservatively taken as
10\%. The resulting uncertainties in the measurements in both the \wenu\ and 
\wmunu\ channels are below 0.2\%. Uncertainties related to the multi-jet background evaluation arise from the statistical precision of the multi-jet templates and uncertainty in the normalisations of the subtracted EW and top-quark contamination. These contributions are propagated through linear 
extrapolations over the isolation variables to the signal region. The related uncertainties in the measurements are evaluated to be 0.7--0.8\% in the \wenu\ channels and not more than 0.2\% in the \wmunu\ channels.

In both the \zmm\ and \zee\ channels, the uncertainty associated with the background subtraction is negligible, since all individual background contributions are below 0.2\% of the selected data sample.

\subsection{Luminosity calibration}

Luminosity measurements in ATLAS are calibrated using dedicated van der Meer scans~\cite{Aaboud:2016hhf}. The analysis of data from the scan performed in \pp collisions at $\sqrt{s}=5.02$~\TeV, which uses the LUCID-2 detector for the baseline luminosity measurements~\cite{Avoni:2018iuv}, yields a relative systematic uncertainty of 1.9\% in the measured luminosity. The largest sources of uncertainty are systematic effects related to the van der Meer scan procedure and the long-term stability of the luminosity calibration

\section{Results \label{sec:results}}
\subsection{Channel combination}

Results of measurements in the electron and muon channels are summarised in Table~\ref{tab:wplusresults} for $W^+$ boson production, Table~\ref{tab:wminusresults} for $W^-$ boson production and
Table~\ref{tab:zresults} for $Z$ boson production. In these tables, the statistical uncertainty is defined from the variance of the background-subtracted number of observed events, and the systematic
uncertainty includes all uncertainty components described above, except for the luminosity uncertainty, which is given separately. The systematic uncertainties coming from lepton efficiency
corrections are measured as a function of lepton ${\eta}$ and \pt, and include a significant statistical component due to the number of \Zboson\ events used to derive the corrections. This
statistical component is substantially reduced for the integrated cross sections compared to the differential ones.

The data tables provided in this paper contain compact summaries of the measurement uncertainties. A complete breakdown of systematic uncertainties and their correlated
components is provided in HEPData~\cite{Maguire:2017ypu}.

\begin{table}
\centering
\caption{Measured fiducial \wplnu\ differential and integrated cross sections for electron and muon channels.\label{tab:wplusresults}}
\begin{tabular}{SS|i{4}i{2}i{2}i{2}|i{4}i{2}i{2}i{2}}
\hline
\hline
                   &                    & \multicolumn{4}{c}{\wpenu} & \multicolumn{4}{c}{\wpmunu} \\
$|{\eta}_{\ell}|^\mathrm{min}$ & $|{\eta}_{\ell}|^\mathrm{max}$ & $\text{d}\sigma/\text{d}|{\eta}_{\ell}|$ & $\delta\sigma_\mathrm{stat}$ & $\delta\sigma_\mathrm{syst}$ & $\delta\sigma_\mathrm{lumi}$ & $\text{d}\sigma/\text{d}|{\eta}_{\ell}|$ & $\delta\sigma_\mathrm{stat}$ & $\delta\sigma_\mathrm{syst}$ & $\delta\sigma_\mathrm{lumi}$ \\
                   &                    & $[\mathrm{pb}]$ & $[\mathrm{pb}]$ & $[\mathrm{pb}]$ & $[\mathrm{pb}]$ & $[\mathrm{pb}]$ & $[\mathrm{pb}]$ & $[\mathrm{pb}]$ & $[\mathrm{pb}]$ \\
\hline
0.00 & 0.21	& 448	& 8	& 10	& 8				& 473	& 9	& 15	& 9 \\
0.21 & 0.42	& 463	& 8	& 10	& 9				& 472	& 8	& 11	& 9 \\
0.42 & 0.63	& 453	& 8	& 10	& 9				& 493	& 8	& 11	& 9 \\
0.63 & 0.84	& 460	& 8	& 10	& 9				& 460	& 9	& 12	& 9 \\
0.84 & 1.05	& 466	& 9	& 11	& 9				& 478	& 9	& 13	& 9 \\
1.05 & 1.37	& 469	& 7	& 10	& 9				& 478	& 6	& 10	& 9 \\
1.37 & 1.52	& {\;\;\;--} 	& {\;\;--} 	& {--} 	& {\;\;--}	& 482	& 9	& 12	& 9 \\
1.52 & 1.74	& 460	& 9	& 14	& 9				& 482	& 7	& 10	& 9 \\
1.74 & 1.95	& 454	& 9	& 14	& 8				& 472	& 8	& 10	& 9 \\
1.95 & 2.18	& 453	& 9	& 14	& 8				& 443	& 7	& 10	& 9 \\
2.18 & 2.50	& 370	& 7	& 14	& 7				& 371	& 7	& 9	& 7 \\
\hline
0.00 & 2.50	& 2243 	& 13	& 27	& 42				& 2303	& 12	& 36	& 44 \\
\hline\hline
\end{tabular}
\end{table}

\begin{table}
\centering
\caption{Measured fiducial \wmlnu\ differential and integrated cross sections for electron and muon channels.\label{tab:wminusresults}}
\begin{tabular}{SS|i{4}i{2}i{2}i{2}|i{4}i{1}i{2}i{2}}
\hline
\hline
                   &                    & \multicolumn{4}{c}{\wmenu} & \multicolumn{4}{c}{\wmmunu} \\
$|{\eta}_{\ell}|^\mathrm{min}$ & $|{\eta}_{\ell}|^\mathrm{max}$ & $\text{d}\sigma/\text{d}|{\eta}_{\ell}|$ & $\delta\sigma_\mathrm{stat}$ & $\delta\sigma_\mathrm{syst}$ & $\delta\sigma_\mathrm{lumi}$ & $\text{d}\sigma/\text{d}|{\eta}_{\ell}|$ & $\delta\sigma_\mathrm{stat}$ & $\delta\sigma_\mathrm{syst}$ & $\delta\sigma_\mathrm{lumi}$ \\
                   &                    & $[\mathrm{pb}]$ & $[\mathrm{pb}]$ & $[\mathrm{pb}]$ & $[\mathrm{pb}]$ & $[\mathrm{pb}]$ & $[\mathrm{pb}]$ & $[\mathrm{pb}]$ & $[\mathrm{pb}]$ \\
\hline
0.00 & 0.21	& 322	& 7	& 7	& 6					& 341	& 8	& 10	& 6 \\
0.21 & 0.42	& 316	& 7	& 7	& 6					& 314	& 7	& 6	& 6 \\
0.42 & 0.63	& 303	& 7	& 7	& 6					& 327	& 7	& 6	& 6 \\
0.63 & 0.84	& 294	& 7	& 7	& 6					& 303	& 7	& 7	& 6 \\
0.84 & 1.05	& 300	& 7	& 7	& 6					& 306	& 7	& 8	& 6 \\
1.05 & 1.37	& 280	& 5	& 6	& 5					& 290	& 5	& 5	& 6 \\
1.37 & 1.52	& {\;\;\;--}	& {\;\;--}	& {\;\;--}	& {\;\;--}	& 276	& 7	& 6	& 5 \\
1.52 & 1.74	& 270	& 7	& 9	& 5					& 272	& 6	& 5	& 5 \\
1.74 & 1.95	& 260	& 7	& 9	& 5					& 245	& 6	& 5	& 5 \\
1.95 & 2.18	& 255	& 7	& 9	& 5					& 253	& 5	& 5	& 5 \\
2.18 & 2.50	& 220	& 6	& 10	& 4					& 219	& 5	& 5	& 4 \\
\hline
0.00 & 2.50	& 1393 	& 10	& 17	& 26					& 1412	& 9	& 22	& 28 \\
\hline\hline
\end{tabular}
\end{table}

\begin{table}
\centering
\caption{Measured fiducial \zll\ differential and integrated cross sections for electron and muon channels.\label{tab:zresults}}
\begin{tabular}{SS|d{3}d{1}d{1}d{1}|d{3}d{1}d{1}d{1}}
\hline
\hline
                   &                    & \multicolumn{4}{c}{\zee} & \multicolumn{4}{c}{\zmm} \\
$|y_{\ell\ell}|^\mathrm{min}$ & $|y_{\ell\ell}|^\mathrm{max}$ & $\text{d}\sigma/\text{d}|y_{\ell\ell}|$ & $\delta\sigma_\mathrm{stat}$ & $\delta\sigma_\mathrm{syst}$ & $\delta\sigma_\mathrm{lumi}$ & $\text{d}\sigma/\text{d}|y_{\ell\ell}|$ & $\delta\sigma_\mathrm{stat}$ & $\delta\sigma_\mathrm{syst}$ & $\delta\sigma_\mathrm{lumi}$ \\
                   &                    & $[\mathrm{pb}]$ & $[\mathrm{pb}]$ & $[\mathrm{pb}]$ & $[\mathrm{pb}]$ & $[\mathrm{pb}]$ & $[\mathrm{pb}]$ & $[\mathrm{pb}]$ & $[\mathrm{pb}]$ \\
\hline
0.0  & 0.5	& 99.9	& 2.5	& 1.6	& 1.9	& 105.2	& 2.4	& 1.1	& 2.0 \\
0.5  & 1.0  	& 100.3	& 2.7	& 1.6	& 1.9	& 101.9	& 2.3	& 1.0	& 1.9 \\
1.0  & 1.5	& 89.2	& 2.7	& 1.4	& 1.7	& 89.8 	& 2.1	& 0.8	& 1.7 \\
1.5  & 2.0  	& 59.6	& 2.4	& 1.2	& 1.1	& 61.0 	& 1.8	& 0.6	& 1.1 \\
2.0  & 2.5	& 19.6	& 1.3	& 0.7	& 0.4	& 20.3 	& 1.2	& 0.2	& 0.4 \\
\hline
0.0  & 2.5	& 369.0	& 5.3	& 4.7	& 6.9	& 377.9	& 4.4	& 3.4	& 7.1 \\
\hline\hline
\end{tabular}
\end{table}

\begin{figure}[!ht]
  \centering
  \subfloat[]{\includegraphics[width=0.49\textwidth]{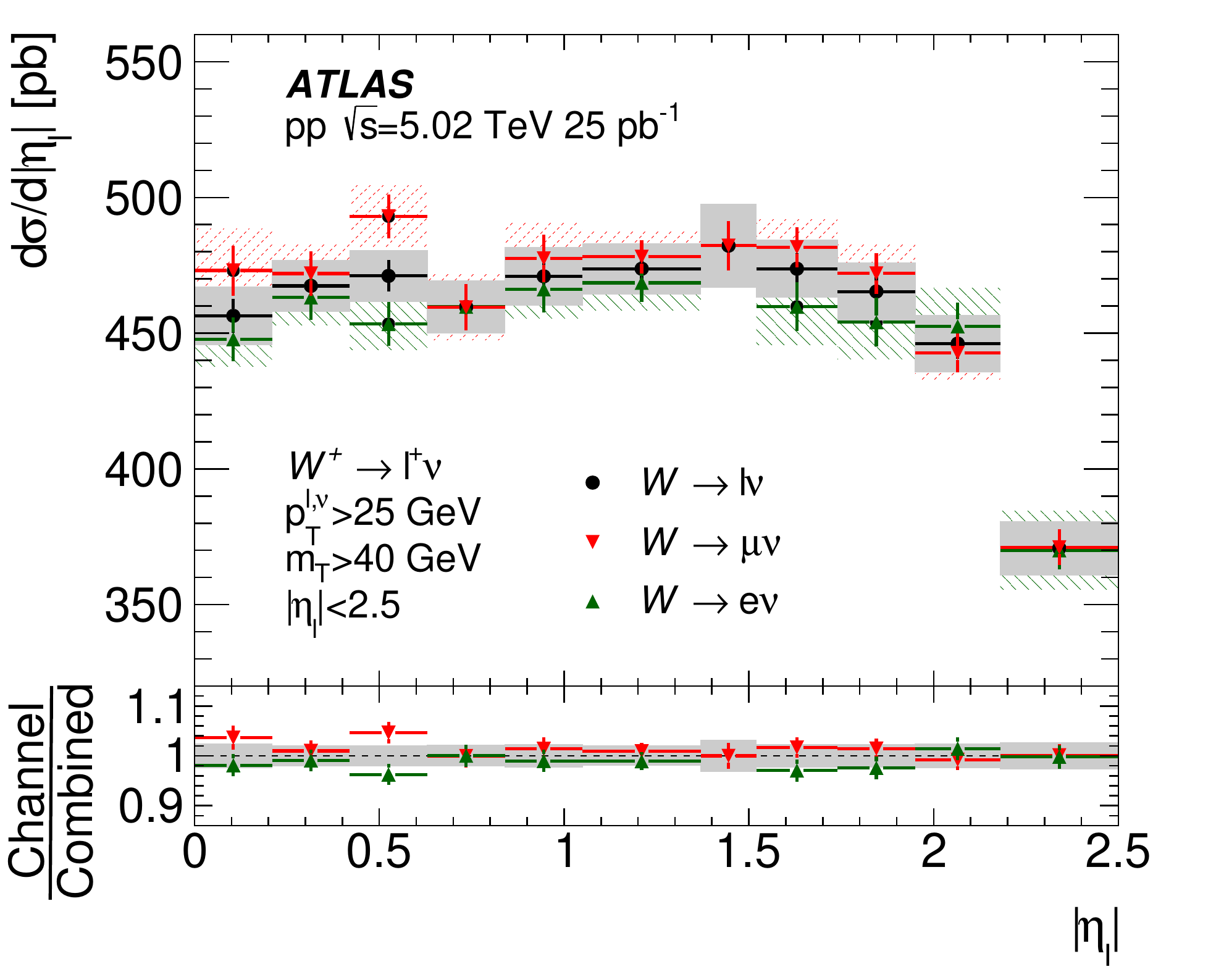}}
  \subfloat[]{\includegraphics[width=0.49\textwidth]{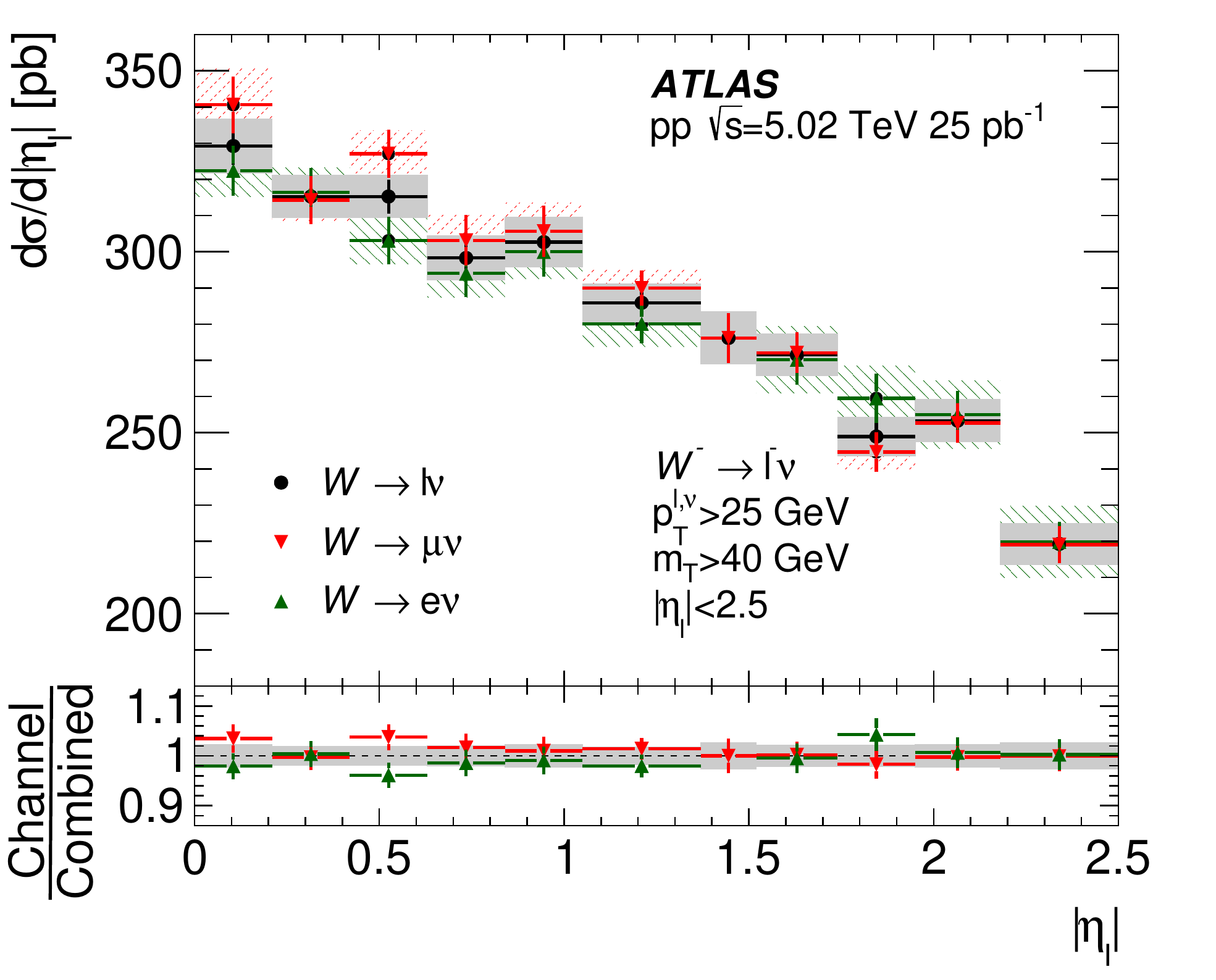}}
  \caption{Differential (a)~$W^+$ and (b)~$W^-$ boson production cross sections as a function of absolute decay lepton pseudorapidity, for the electron, muon and combined results. Statistical and systematic errors are shown as corresponding bars and shaded bands. The luminosity uncertainty is not included. The lower panel shows the ratio of channels to the combined differential cross section in each bin. In the lower panel, error bars represent statistical uncertainties in the ratio, while the shaded band represents systematic uncertainties in the combined differential cross sections.}
  \label{fig:comb1}
\end{figure}

\begin{figure}[!ht]
  \centering
  \includegraphics[width=0.49\textwidth]{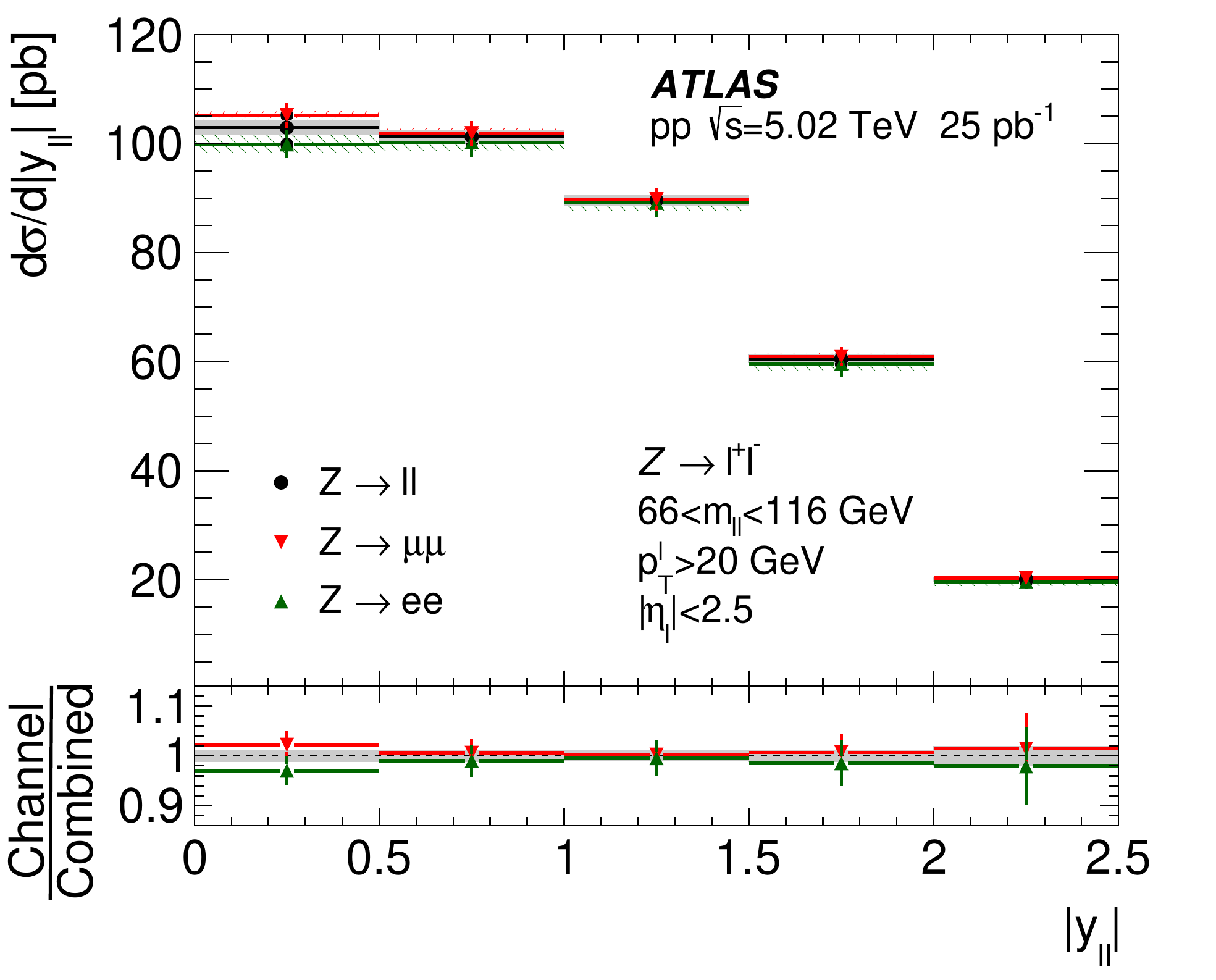}
  \caption{Differential $Z$ boson production cross section as a function of absolute lepton-pair rapidity, for the electron, muon and combined results. Statistical and systematic errors are shown as corresponding bars and shaded bands.  The luminosity uncertainty is not included. The lower panel shows the ratio of channels to the combined differential cross section in each bin. In the lower panel, error bars represent statistical uncertainties in the ratio, while the shaded band represents systematic uncertainties in the combined differential cross sections.}
  \label{fig:comb2}
\end{figure}

The electron and muon channel measurements are combined using the Best Linear Unbiased Estimate (BLUE) method~\cite{Valassi:2003mu}, accounting for the correlations of the systematic uncertainties across the
channels and measurement bins. The $|\eta_\ell|$ and $|\yll|$ distributions for the electron channel, muon channel and combined results are shown in
Figures~\ref{fig:comb1} and~\ref{fig:comb2} for $W$ and $Z$ bosons, respectively, and the results are listed in Tables~\ref{tab:wpcomb}--\ref{tab:zcomb}. In the interval $1.37<|\eta_\ell|<1.52$, only the muon channel measurements for $W$ boson production are used. The combination yields $\chi^2/$d.o.f$=19.3/10$ for the $W^+$ boson results, $\chi^2/$d.o.f$=15.1/10$ for the $W^-$ boson results, and
$\chi^2/$d.o.f$=3.0/5$ for the $Z$ boson results. A simultaneous combination of all measurements, accounting for the correlation of the experimental systematic uncertainties between the $W$ and $Z$
measurement results for a given lepton flavour, gives $\chi^2/$d.o.f~$=37.5/25$, corresponding to a probability of 5.2\%. In view of this remaining discrepancy and of the general trend of the muon
channel cross sections to be higher than the electron channel ones, the systematic uncertainties in the efficiency corrections are scaled such that  $\chi^2/$d.o.f=1; the correction
uncertainties are scaled by a common factor, preserving the uncertainty correlations as a function of lepton \pt\ and $\eta$ for this source.
Tables~\ref{tab:wpcomb}--\ref{tab:zcomb} include this scaling. The measured ratio of fiducial $W^+$ and $W^-$ production cross sections, as well as ratios of fiducial \Wpm\ and \Z\ production cross sections, are summarised in Table~\ref{tab:rfid}.

\begin{table}
  \centering
\caption{Combined fiducial \wplnu\ differential and integrated cross sections.\label{tab:wpcomb}}
  \begin{tabular}{SS|i{4}i{1}i{2}i{2}}
    \hline
\hline
                   &                    & \multicolumn{4}{c}{\wplnu} \\
$|{\eta}_{\ell}|^\mathrm{min}$ & $|{\eta}_{\ell}|^\mathrm{max}$ & $\text{d}\sigma/\text{d}|{\eta}_{\ell}|\,[\mathrm{pb}]$ & $\delta\sigma_\mathrm{stat}\,[\mathrm{pb}]$& $\delta\sigma_\mathrm{syst}\,[\mathrm{pb}]$ & $\delta\sigma_\mathrm{lumi}\,[\mathrm{pb}]$  \\
\hline
 0.00 &    0.21 &  456 &               6 &              11 &               9 \\
 0.21 &    0.42 &  467 &               6 &               9 &               9 \\
 0.42 &    0.63 &  471 &               6 &               9 &               9 \\
 0.63 &    0.84 &  460 &               6 &              10 &               9 \\
 0.84 &    1.05 &  471 &               6 &              11 &               9 \\
 1.05 &    1.37 &  474 &               5 &               9 &               9 \\
 1.37 &    1.52 &  482 &               9 &              15 &               9 \\
 1.52 &    1.74 &  474 &               6 &              11 &               9 \\
 1.74 &    1.95 &  465 &               6 &              11 &               9 \\
 1.95 &    2.18 &  446 &               6 &              10 &               9 \\
 2.18 &    2.50 &  371 &               5 &              10 &               7 \\
\hline
 0.00 &    2.50 & 2266 &               9 &              29 &              43 \\
\hline\hline
\end{tabular}
\end{table}

\begin{table}
  \centering
\caption{Combined fiducial \wmlnu\ differential and integrated cross sections.\label{tab:wmcomb}}
  \begin{tabular}{SS|i{4}i{1}i{2}i{2}}
    \hline
\hline
                   &                    & \multicolumn{4}{c}{\wmlnu} \\
$|{\eta}_{\ell}|^\mathrm{min}$ & $|{\eta}_{\ell}|^\mathrm{max}$ & $\text{d}\sigma/\text{d}|{\eta}_{\ell}|\,[\mathrm{pb}]$ & $\delta\sigma_\mathrm{stat}\,[\mathrm{pb}]$& $\delta\sigma_\mathrm{syst}\,[\mathrm{pb}]$ & $\delta\sigma_\mathrm{lumi}\,[\mathrm{pb}]$  \\
\hline
   0.00 &    0.21 &    329 &               5 &               8 &               6 \\    
   0.21 &    0.42 &    315 &               5 &               6 &               6 \\    
   0.42 &    0.63 &    315 &               5 &               6 &               6 \\    
   0.63 &    0.84 &    298 &               5 &               6 &               6 \\    
   0.84 &    1.05 &    303 &               5 &               7 &               6 \\    
   1.05 &    1.37 &    286 &               4 &               5 &               6 \\    
   1.37 &    1.52 &    276 &               7 &               7 &               5 \\    
   1.52 &    1.74 &    272 &               4 &               6 &               5 \\    
   1.74 &    1.95 &    249 &               4 &               5 &               5 \\    
   1.95 &    2.18 &    253 &               4 &               6 &               5 \\    
   2.18 &    2.50 &    219 &               4 &               6 &               4 \\    
\hline
   0.00 &    2.50 &   1401 &               7 &              18 &              27 \\
\hline\hline
\end{tabular}
\end{table}

\begin{table}
\centering
\caption{Combined fiducial \zll\ differential and integrated cross sections.\label{tab:zcomb}}
\begin{tabular}{SS|d{3}d{1}d{1}d{1}}
\hline
\hline
                   &                    & \multicolumn{3}{c}{\zll} \\
$|y_{\ell\ell}|^\mathrm{min}$ & $|y_{\ell\ell}|^\mathrm{max}$ & $\text{d}\sigma/\text{d}|y_{\ell\ell}|\,[\mathrm{pb}]$ & $\delta\sigma_\mathrm{stat}\,[\mathrm{pb}]$& $\delta\sigma_\mathrm{syst}\,[\mathrm{pb}]$ & $\delta\sigma_\mathrm{lumi}\,[\mathrm{pb}]$ \\
\hline
0.0 &     0.5 &   103.0 &             1.7 &             1.2 &             1.9 \\ 
0.5 &     1.0 &   101.3 &             1.8 &             1.1 &             1.9 \\ 
1.0 &     1.5 &    89.6 &             1.7 &             0.9 &             1.7 \\ 
1.5 &     2.0 &    60.5 &             1.4 &             0.7 &             1.1 \\ 
2.0 &     2.5 &    20.0 &             0.9 &             0.4 &             0.4 \\ 
\hline
0.0 &     2.5 &   374.5 &             3.4 &             3.6 &             7.0 \\

\hline\hline
\end{tabular}
\end{table}

\begin{table}
\centering
\caption{Ratios of integrated $W$ and $Z$ production cross sections.\label{tab:rfid}}
\begin{tabular}{c|c}
\hline
\hline
$\RFid_{W^+/W^-}$ & 1.617 $\pm$ 0.012 (stat) $\pm$ 0.003 (syst) \\
$\RFid_{W/Z}$     & 9.81 $\pm$ 0.13 (stat) $\pm$ 0.01 (syst) \\
$\RFid_{W^+/Z}$   & 6.06 $\pm$ 0.08 (stat) $\pm$ 0.01 (syst) \\
$\RFid_{W^-/Z}$   & 3.75 $\pm$ 0.05 (stat) $\pm$ 0.01 (syst) \\
\hline\hline
\end{tabular}
\end{table}

\begin{figure}[!ht]
  \centering
  \includegraphics[width=0.49\textwidth]{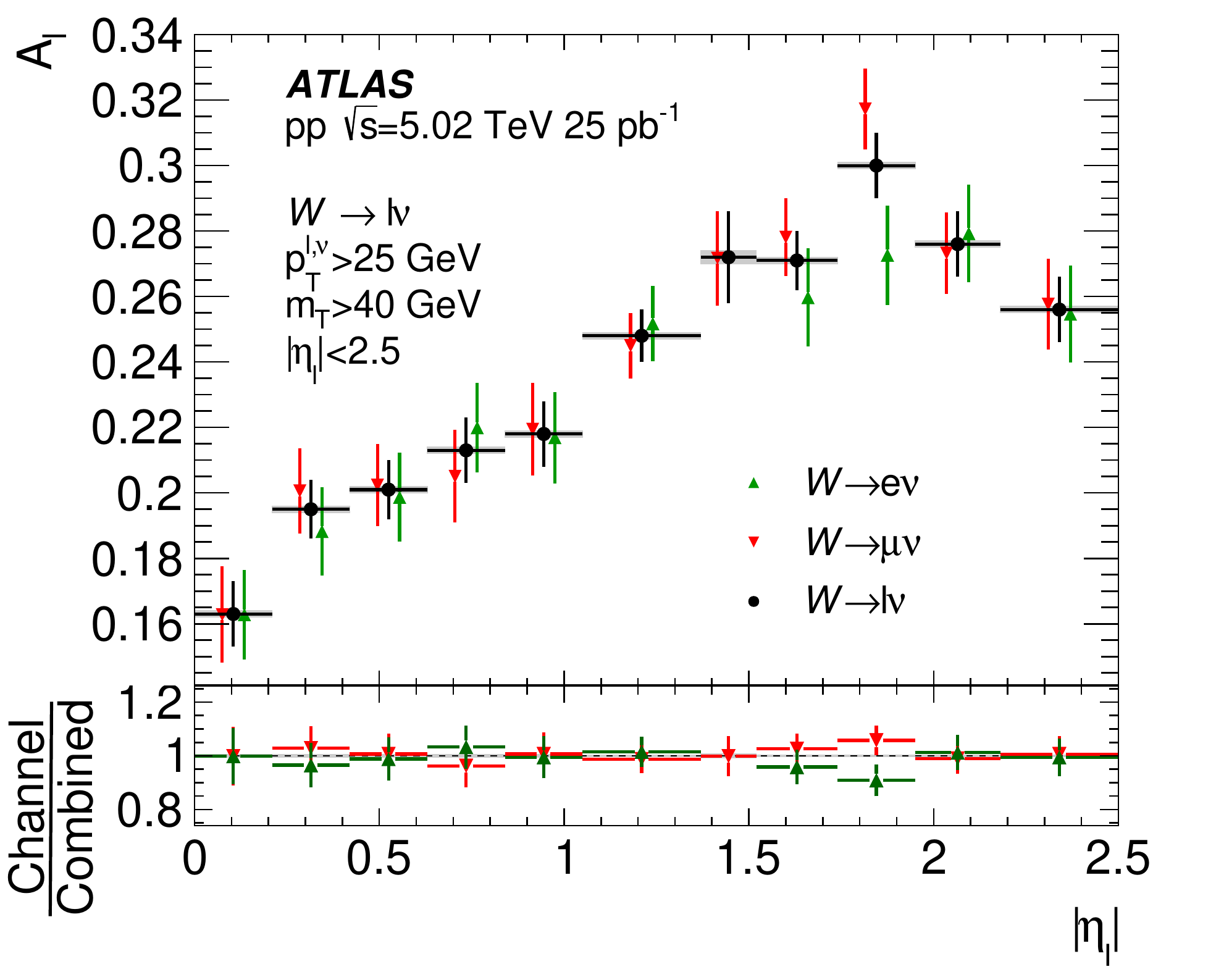}
  \caption{Charge asymmetry for $W$ bosons as a function of absolute decay lepton pseudorapidity, for the electron, muon and combined results. Statistical and systematic errors are shown as corresponding bars and shaded bands (not visible for most points). The lower panel shows the ratio of channels to the combined charge asymmetry in each bin. In the lower panel, error bars represent statistical uncertainties in the ratio, while the shaded band represents systematic uncertainties in the combined charge asymmetry.}
  \label{fig:comb3}
\end{figure}

The measurements of differential $W^+$ and $W^-$ production cross sections allow the extraction of the $W$ boson charge asymmetry, as a function of the absolute pseudorapidity of the decay lepton:
\begin{equation}
\nonumber 
  A_\ell(|\eta_\ell|) = \frac{\text{d}\sigma_{W^+} / \text{d}|\eta_\ell| - \text{d}\sigma_{W^-} / \text{d}|\eta_\ell|}{\text{d}\sigma_{W^+} / \text{d}|\eta_\ell| + \text{d}\sigma_{W^-} / \text{d}|\eta_\ell|}.
\label{eq:Al} 
\end{equation}
Uncertainties in $A_\ell$ are calculated considering all sources of correlated and uncorrelated systematic uncertainties 
in the differential cross sections. The resulting dependence of $A_\ell$ on $|\eta_\ell|$ measured in the electron and muon channels is presented in Figure~\ref{fig:comb3} together with the combined values, while the combined results are summarised with the corresponding uncertainties in Table~\ref{tab:wasym}. Good agreement between the two channels is found.

\begin{table}
  \centering
  \caption{Charge asymmetry for $W$ bosons as a function of absolute pseudorapidity of the decay lepton.\label{tab:wasym}}
  \begin{tabular}{cc|cccc}
    \hline
    \hline
    $|{\eta}_{\ell}|^\mathrm{min}$ & $|{\eta}_{\ell}|^\mathrm{max}$ & $A_\ell$ & $\delta A_\mathrm{stat}$ & $\delta A_\mathrm{syst}$ \\
    \hline
    0.00 &    0.21 &        0.163 &           0.010 &           0.001 \\
    0.21 &    0.42 &        0.195 &           0.009 &           0.001 \\
    0.42 &    0.63 &        0.201 &           0.009 &           0.001 \\
    0.63 &    0.84 &        0.213 &           0.010 &           0.001 \\
    0.84 &    1.05 &        0.218 &           0.010 &           0.001 \\
    1.05 &    1.37 &        0.248 &           0.008 &           0.001 \\
    1.37 &    1.52 &        0.272 &           0.014 &           0.002 \\
    1.52 &    1.74 &        0.271 &           0.009 &           0.001 \\
    1.74 &    1.95 &        0.300 &           0.010 &           0.001 \\
    1.95 &    2.18 &        0.276 &           0.010 &           0.001 \\
    2.18 &    2.50 &        0.256 &           0.010 &           0.001 \\
    \hline\hline
  \end{tabular}
\end{table}

\subsection{Comparison with theoretical predictions}

The measured cross sections are compared with theoretical predictions obtained using a modified version of \textsc{DYNNLO}~1.5~\cite{Catani:2007vq,Catani:2009sm} optimised for speed of computation. The
calculation is performed at $O(\alphas^2)$ in QCD and at leading order in the EW theory, with parameters set according to the $G_\mu$ scheme~\cite{Hollik:1988ii}. The input parameters (the
Fermi constant $G_\text{F}$,
the masses and widths of \Wboson\ and \Zboson\ bosons, and the CKM matrix elements) are taken from Ref.~\cite{Agashe:2014kda}. 
The \textsc{DYNNLO} predictions are calculated using the NNLO PDF sets from \textsc{CT14nnlo}~\cite{Dulat:2015mca}, \textsc{NNPDF3.1}~\cite{Ball:2017nwa}, \textsc{MMHT14nnlo68cl}~\cite{Harland-Lang:2014zoa}, \textsc{HERAPDF2.0}~\cite{Abramowicz:2015mha} and \textsc{ABMP16}~\cite{Alekhin:2017kpj}.
All considered PDF sets except \textsc{HERAPDF2.0} are evaluated from global fits which include to varying extents the LHC measurements of $W/Z$ boson, Drell-Yan, top-quark and inclusive jet production.
The renormalisation and factorisation scales, respectively denoted as $\muR$ and $\muF$, are set equal to the decay lepton-pair invariant mass, $m_{\ell\nu}$ or $\mll$.

Uncertainties in these predictions are derived as follows. PDF uncertainties are evaluated from the variations of the NNLO PDFs (the PDF uncertainties of \textsc{CT14nnlo} are rescaled from 90\% confidence level to
68\% confidence level). Scale uncertainties are defined by the envelope of the variations obtained by changing $\muR$ and $\muF$ by a factor of two with respect to their nominal values and imposing $0.5\leq \muR/\muF \leq2$. The uncertainty induced by the strong coupling constant is estimated by varying $\alphas$ by $\pm 0.001$ around the central value of $\alphas(m_Z) = 0.118$,
following the prescription of Ref.~\cite{Dulat:2015mca}; the effect of these variations is estimated by comparing the \textsc{CT14nnlo\_as\_0117} and \textsc{CT14nnlo\_as\_0119} PDF sets to \textsc{CT14nnlo}. Finally, intrinsic
limitations of the NNLO calculations for fiducial cross-section predictions lead to systematic differences between results from different programs, as explained in
Ref.~\cite{Aaboud:2016zpd}. Therefore, an additional uncertainty of 0.7\%, estimated from a comparison of predictions calculated with \textsc{Fewz}~3.1 and \textsc{DYNNLO}, is assigned. Theory uncertainties are dominated by our knowledge of the proton PDFs.

The uncertainty of the LHC proton beam energy is estimated to be 0.1\%~\cite{Todesco:2017nnk} and induces typically an uncertainty of 0.1\% in the cross-section predictions, which is negligible compared to other theoretical uncertainties discussed above.

\begin{figure}[!ht]
  \centering
  \subfloat[]{\includegraphics[width=0.49\textwidth]{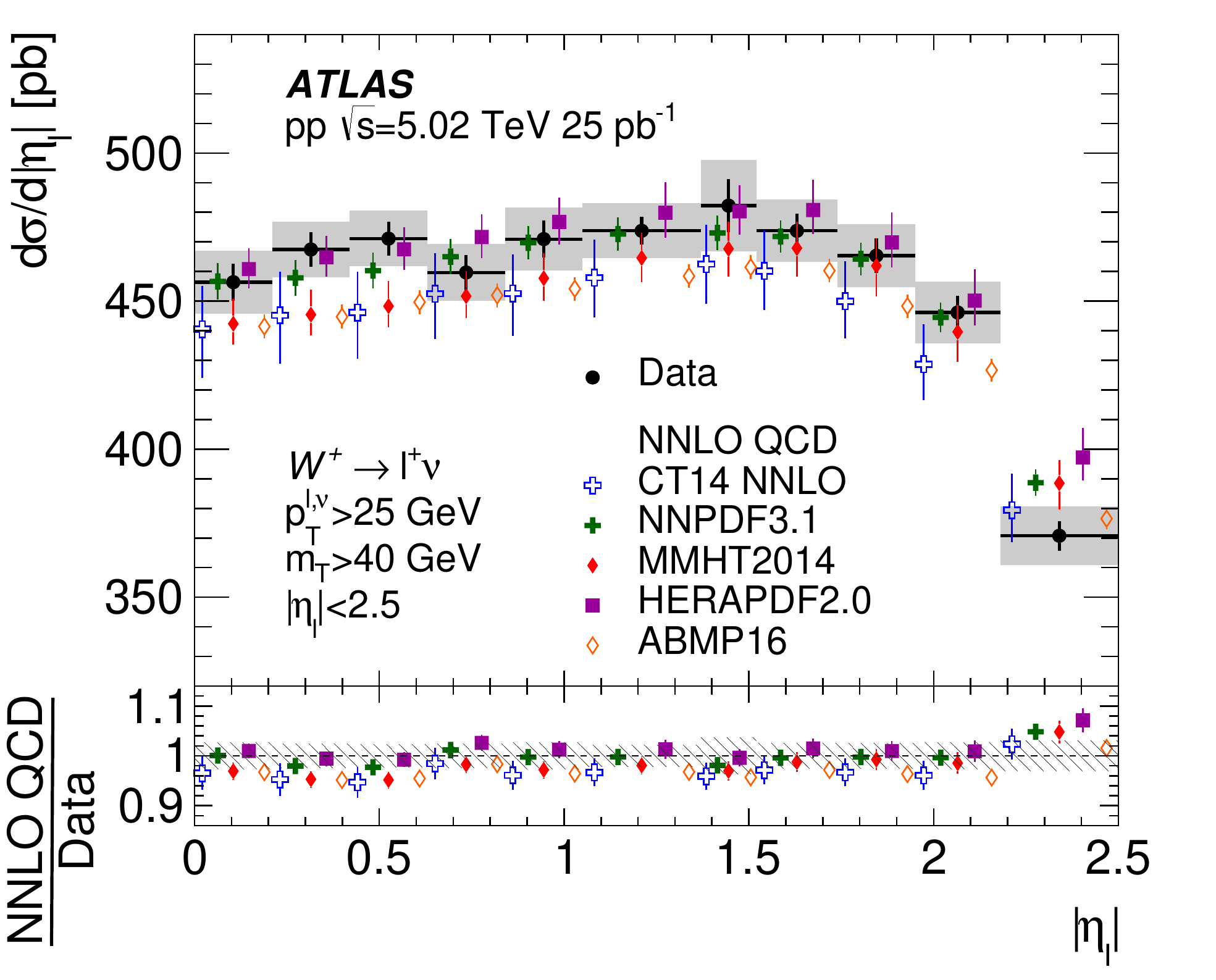}}
  \subfloat[]{\includegraphics[width=0.49\textwidth]{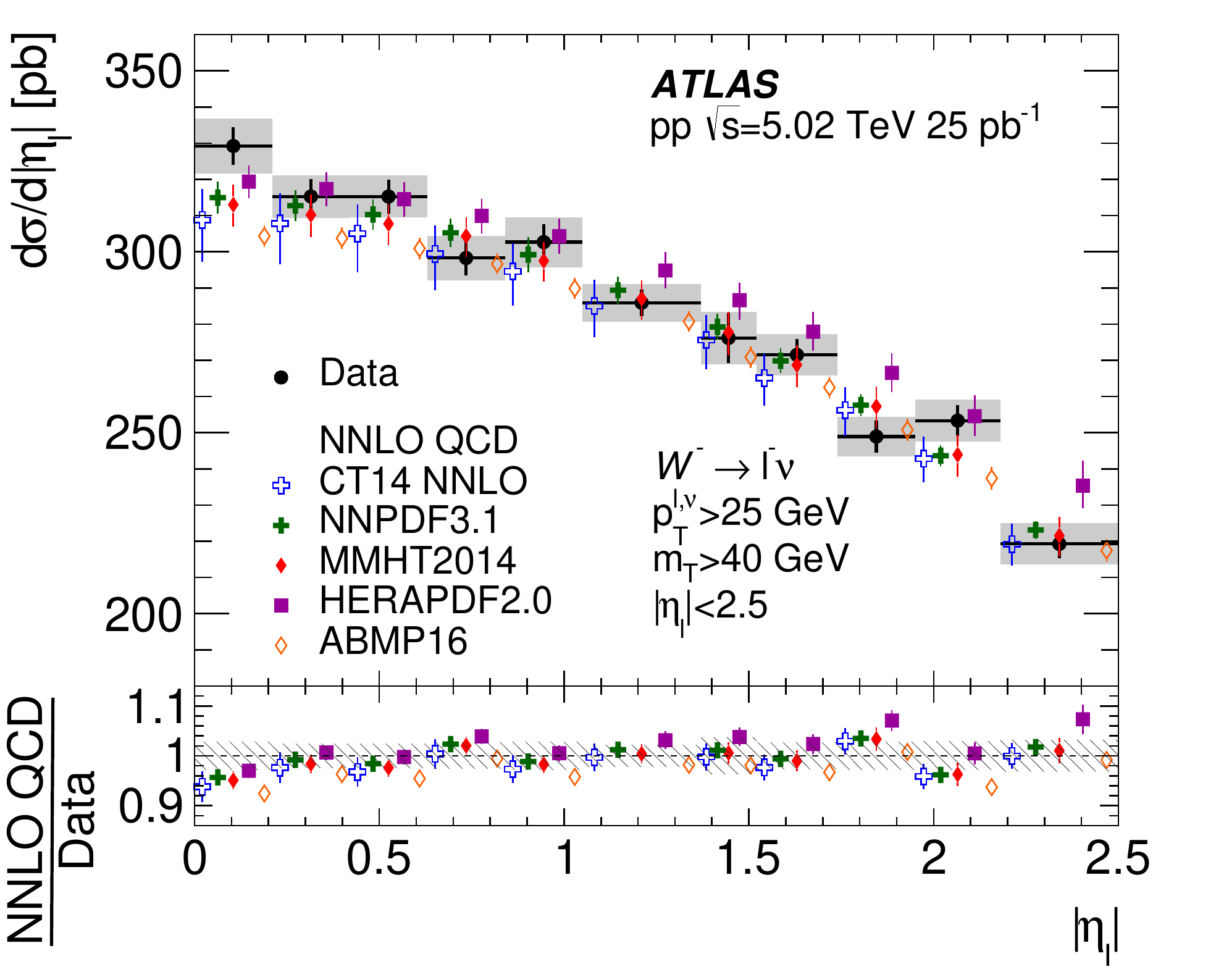}}
  \caption{Differential cross sections for (a)~$W^+$ and (b)~$W^-$ boson production as a function of absolute decay lepton pseudorapidity compared with theoretical predictions.  Statistical and systematic errors are shown as corresponding bars and shaded bands on the data points. The luminosity uncertainty is not included. Only the dominant uncertainty~(PDF) is displayed for the theory. The lower panel shows the ratio of predictions to the measured differential cross section in each bin, and the shaded band shows the sum in quadrature of statistical and systematic uncertainties of the data.
  \label{fig:res-w-diff}}
\end{figure}

\begin{figure}[!ht]
  \centering
  \includegraphics[width=0.49\textwidth]{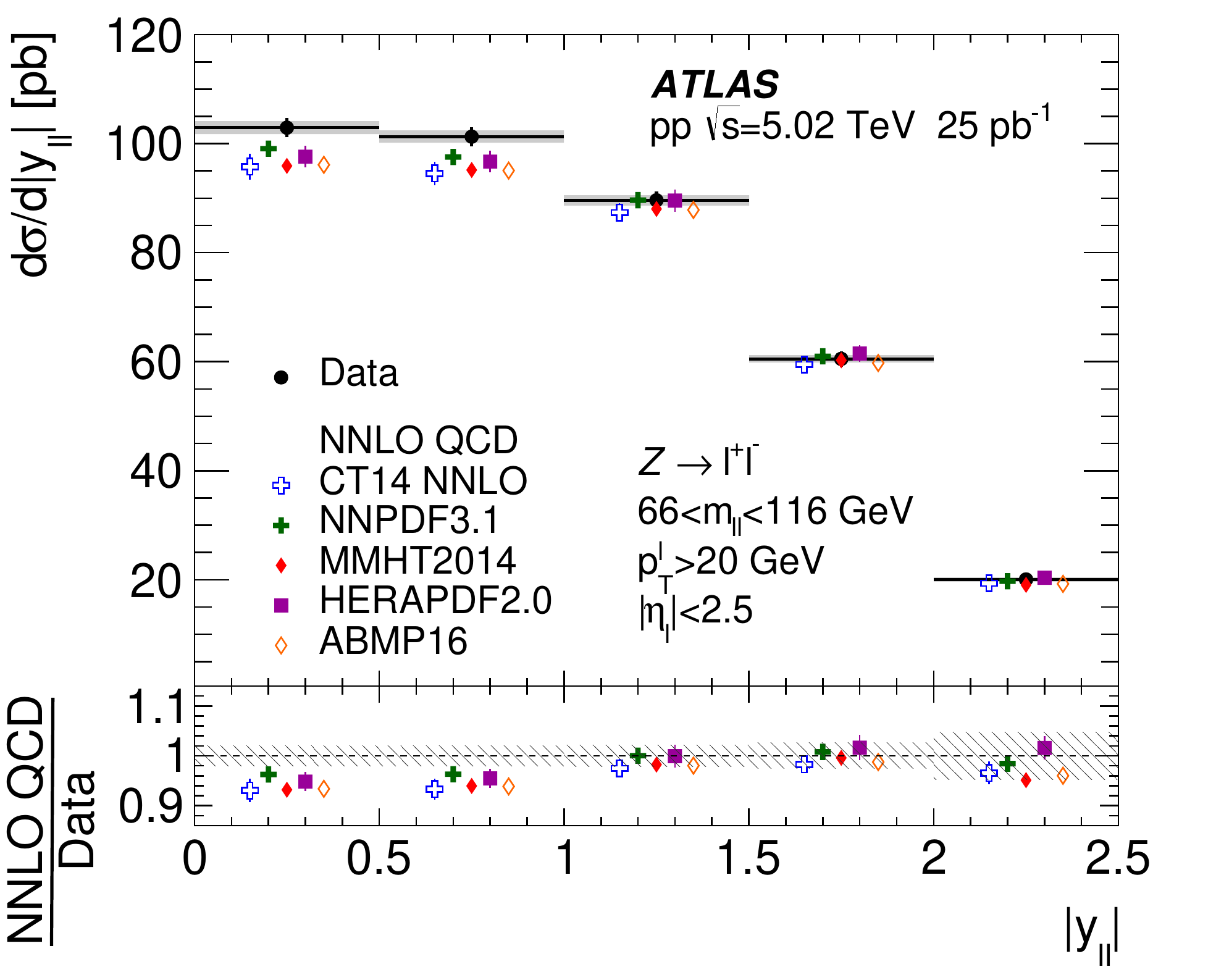}
  \caption{Differential cross section for $Z$ boson production as a function of absolute lepton-pair rapidity compared
with theoretical predictions.  Statistical and systematic errors are shown as corresponding bars and shaded bands on the data points. The luminosity uncertainty is not included. Only the dominant uncertainty~(PDF) is displayed for the theory. The lower panel shows the ratio of predictions to the measured differential cross section in each bin, and the shaded band shows the sum in quadrature of statistical and systematic uncertainties of the data.
  \label{fig:res-z-diff}}
\end{figure}

Differential cross sections for $W$ and $Z$ boson production are shown in Figures~\ref{fig:res-w-diff} and \ref{fig:res-z-diff} as a function of $|\eta_{\ell}|$ and $|y_{\ell\ell}|$, respectively. The
cross sections are compared for the combined measurement and theoretical predictions calculated with the \textsc{CT14nnlo}, \textsc{NNPDF3.1}, \textsc{MMHT14nnlo68cl}, \textsc{HERAPDF2.0} and \textsc{ABMP16} PDF sets, with uncertainties assigned as described above.
A comparison of the differential cross sections shows that the predictions obtained with the \textsc{NNPDF3.1} PDF set are in good agreement with the measured values, mainly because the \textsc{NNPDF3.1} global fit includes high precision LHC measurements of $W/Z$ boson production~\cite{Aaboud:2016btc, CMS-SMP-14-022}.
On the other hand, the predictions obtained with other recent PDF sets systematically deviate from the measured values. These deviations are largest for $W^{+}$ boson production and at central rapidity for $Z$ boson production.

\begin{figure}[!ht]
  \centering
  \includegraphics[width=0.49\textwidth]{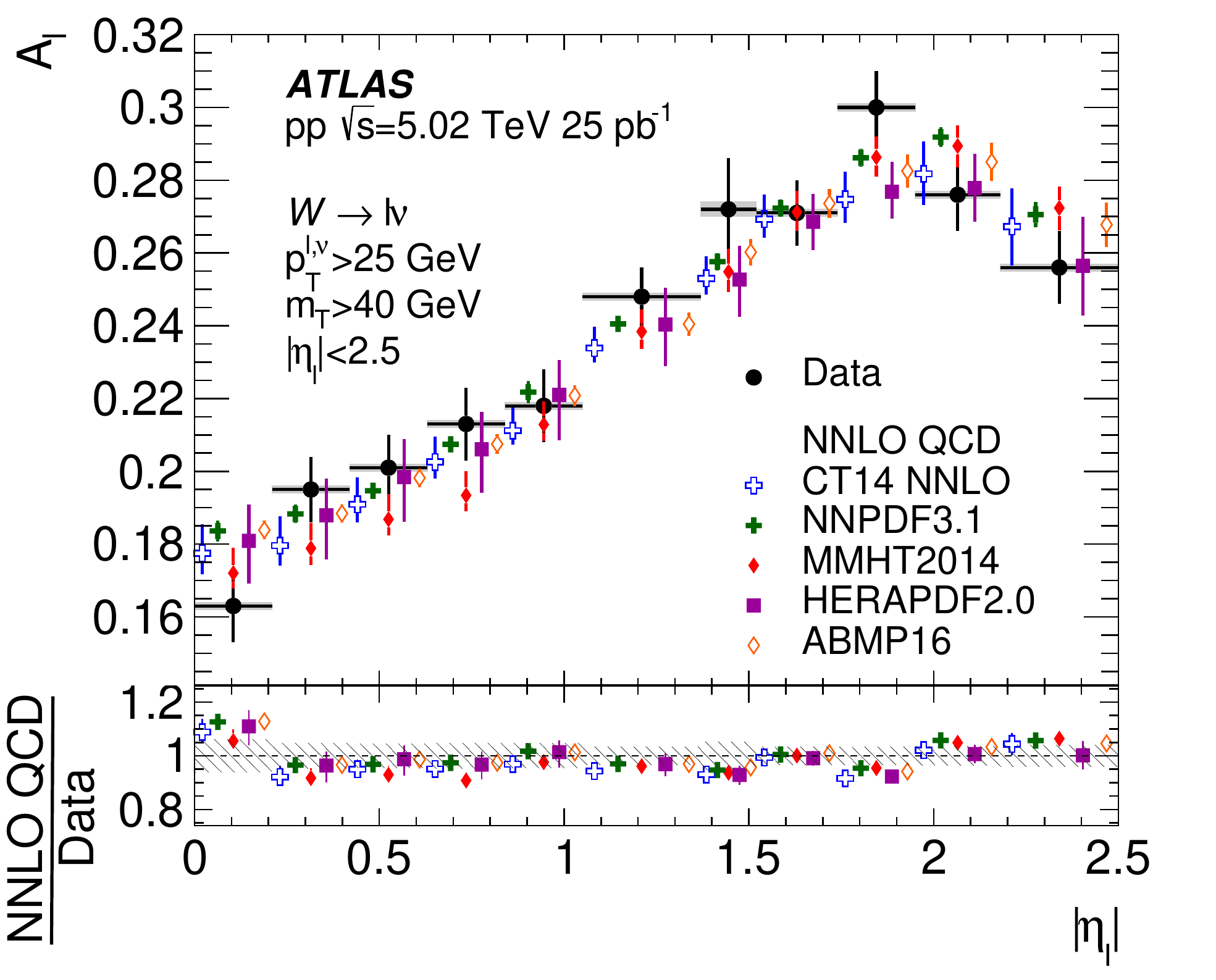}
  \caption{Charge asymmetry for $W$ bosons as a function of absolute decay lepton pseudorapidity compared with theoretical predictions. Statistical and systematic errors are shown as corresponding bars and shaded bands on the data points. Only the dominant uncertainty~(PDF) is displayed for the theory. The lower panel shows the ratio of predictions to the measured differential cross section in each bin, and the shaded band shows the sum in quadrature of statistical and systematic uncertainties of the data.}
  \label{fig:res-w-asym}
\end{figure}

The measured lepton charge asymmetry for $W$ bosons shown in Figure~\ref{fig:res-w-asym} is compared with predictions calculated with the PDF sets mentioned previously. In most of the $|\eta_\ell|$ range considered, the predictions from all PDF sets tend to underestimate the measured asymmetry by a few percent.

\clearpage
\section{Summary \label{sec:summary}}
Fiducial cross sections are reported for inclusive $W^+$, $W^-$ and $Z$ boson production in \pp collisions at the centre-of-mass energy $\sqrt{s}=5.02$~\TeV. The measurement is based on data taken by the ATLAS detector at the LHC corresponding to an integrated luminosity of $25.0\;\mathrm{pb}^{-1}$.  Cross sections are reported in the electron and muon decay channels, integrated over the fiducial regions and differentially. The fiducial region is defined using lepton kinematics and detector acceptance. The differential cross sections for \wlnu\ boson production are measured as a function of absolute lepton pseudorapidity while for \zll\ bosons they are reported as a function of absolute dilepton rapidity in the mass window $66<\mll<116$~\GeV. For \Wpm\ bosons the decay lepton charge asymmetry as a function of absolute lepton pseudorapidity is also measured.

The electron and muon channel results are found to agree within the measurement precision, and are therefore combined considering all sources of correlated and uncorrelated uncertainties. The combined fiducial $W^+$, $W^-$, and $Z$ cross sections are measured with a precision of 1.2--1.7\%, excluding the luminosity uncertainty. Both the integrated and differential cross sections are compared with next-to-next-to-leading-order QCD calculations using various PDF sets. A comparison of the differential cross sections shows 1--2$\sigma$ deviations from the predictions obtained with many of the recent PDF sets.

These results provide the first measurement of $W^\pm$ and $Z$ boson production cross sections at the centre-of-mass energy $\sqrt{s}=5.02$~TeV and complement 
previous measurements at $\sqrt{s}=7$, 8 and 13~TeV.
They constitute a reference for measurements of $W^\pm$ and $Z$ boson production in heavy-ion collisions 
collected at $\sqrt{s_{\mathrm{NN}}}=5.02$~\TeV\ by the LHC experiments.

\clearpage
\section*{Acknowledgements}

We thank CERN for the very successful operation of the LHC, as well as the
support staff from our institutions without whom ATLAS could not be
operated efficiently.

We acknowledge the support of ANPCyT, Argentina; YerPhI, Armenia; ARC, Australia; BMWFW and FWF, Austria; ANAS, Azerbaijan; SSTC, Belarus; CNPq and FAPESP, Brazil; NSERC, NRC and CFI, Canada; CERN; CONICYT, Chile; CAS, MOST and NSFC, China; COLCIENCIAS, Colombia; MSMT CR, MPO CR and VSC CR, Czech Republic; DNRF and DNSRC, Denmark; IN2P3-CNRS, CEA-DRF/IRFU, France; SRNSFG, Georgia; BMBF, HGF, and MPG, Germany; GSRT, Greece; RGC, Hong Kong SAR, China; ISF and Benoziyo Center, Israel; INFN, Italy; MEXT and JSPS, Japan; CNRST, Morocco; NWO, Netherlands; RCN, Norway; MNiSW and NCN, Poland; FCT, Portugal; MNE/IFA, Romania; MES of Russia and NRC KI, Russian Federation; JINR; MESTD, Serbia; MSSR, Slovakia; ARRS and MIZ\v{S}, Slovenia; DST/NRF, South Africa; MINECO, Spain; SRC and Wallenberg Foundation, Sweden; SERI, SNSF and Cantons of Bern and Geneva, Switzerland; MOST, Taiwan; TAEK, Turkey; STFC, United Kingdom; DOE and NSF, United States of America. In addition, individual groups and members have received support from BCKDF, CANARIE, CRC and Compute Canada, Canada; COST, ERC, ERDF, Horizon 2020, and Marie Sk{\l}odowska-Curie Actions, European Union; Investissements d' Avenir Labex and Idex, ANR, France; DFG and AvH Foundation, Germany; Herakleitos, Thales and Aristeia programmes co-financed by EU-ESF and the Greek NSRF, Greece; BSF-NSF and GIF, Israel; CERCA Programme Generalitat de Catalunya, Spain; The Royal Society and Leverhulme Trust, United Kingdom. 

The crucial computing support from all WLCG partners is acknowledged gratefully, in particular from CERN, the ATLAS Tier-1 facilities at TRIUMF (Canada), NDGF (Denmark, Norway, Sweden), CC-IN2P3 (France), KIT/GridKA (Germany), INFN-CNAF (Italy), NL-T1 (Netherlands), PIC (Spain), ASGC (Taiwan), RAL (UK) and BNL (USA), the Tier-2 facilities worldwide and large non-WLCG resource providers. Major contributors of computing resources are listed in Ref.~\cite{ATL-GEN-PUB-2016-002}.

\printbibliography

\clearpage
 
\begin{flushleft}
{\Large The ATLAS Collaboration}

\bigskip

M.~Aaboud$^\textrm{\scriptsize 34d}$,    
G.~Aad$^\textrm{\scriptsize 99}$,    
B.~Abbott$^\textrm{\scriptsize 125}$,    
O.~Abdinov$^\textrm{\scriptsize 13,*}$,    
B.~Abeloos$^\textrm{\scriptsize 129}$,    
D.K.~Abhayasinghe$^\textrm{\scriptsize 91}$,    
S.H.~Abidi$^\textrm{\scriptsize 164}$,    
O.S.~AbouZeid$^\textrm{\scriptsize 39}$,    
N.L.~Abraham$^\textrm{\scriptsize 153}$,    
H.~Abramowicz$^\textrm{\scriptsize 158}$
% ...
\bigskip
\\

% ...
$^{13}$Institute of Physics, Azerbaijan Academy of Sciences, Baku; Azerbaijan.\\
% ...
$^{34}$$^{(a)}$Facult\'e des Sciences Ain Chock, R\'eseau Universitaire de Physique des Hautes Energies - Universit\'e Hassan II, Casablanca;$^{(b)}$Centre National de l'Energie des Sciences Techniques Nucleaires (CNESTEN), Rabat;$^{(c)}$Facult\'e des Sciences Semlalia, Universit\'e Cadi Ayyad, LPHEA-Marrakech;$^{(d)}$Facult\'e des Sciences, Universit\'e Mohamed Premier and LPTPM, Oujda;$^{(e)}$Facult\'e des sciences, Universit\'e Mohammed V, Rabat; Morocco.\\
% ...
$^{39}$Niels Bohr Institute, University of Copenhagen, Copenhagen; Denmark.\\
% ...
$^{91}$Department of Physics, Royal Holloway University of London, Egham; United Kingdom.\\
% ...
$^{99}$CPPM, Aix-Marseille Universit\'e, CNRS/IN2P3, Marseille; France.\\
% ...
$^{125}$Homer L. Dodge Department of Physics and Astronomy, University of Oklahoma, Norman OK; United States of America.\\
% ...
$^{129}$LAL, Universit\'e Paris-Sud, CNRS/IN2P3, Universit\'e Paris-Saclay, Orsay; France.\\
% ...
$^{153}$Department of Physics and Astronomy, University of Sussex, Brighton; United Kingdom.\\
% ...
$^{158}$Raymond and Beverly Sackler School of Physics and Astronomy, Tel Aviv University, Tel Aviv; Israel.\\
% ...
$^{164}$Department of Physics, University of Toronto, Toronto ON; Canada.\\
% ...
$^{a}$ Also at Borough of Manhattan Community College, City University of New York, NY; United States of America.\\
$^{b}$ Also at California State University, East Bay; United States of America.\\
$^{c}$ Also at Centre for High Performance Computing, CSIR Campus, Rosebank, Cape Town; South Africa.\\
$^{d}$ Also at CERN, Geneva; Switzerland.\\
$^{e}$ Also at CPPM, Aix-Marseille Universit\'e, CNRS/IN2P3, Marseille; France.\\
% ...
$^{*}$ Deceased

\end{flushleft}


\end{document}